\documentclass[a4paper,12pt]{article}

\usepackage{jheppub} 
\usepackage[utf8]{inputenc}

\usepackage{dsfont}

\usepackage{float}
\usepackage{ulem}
\usepackage{bm}
\usepackage{cancel}
\usepackage{enumitem}

\usepackage{comment}

\usepackage{color}
\usepackage{graphicx}

\usepackage{shuffle}

\renewcommand{\thefootnote}{\fnsymbol{footnote}}
\renewcommand{\thanks}[1]{\footnote{#1}}

\newcommand{\bea}{\begin{eqnarray}}
\newcommand{\eea}{\end{eqnarray}}
\newcommand{\ee}{\end{equation}}
\newcommand{\be}{\begin{equation}}

\newcommand{\no}{\nonumber}

\def\eqn#1{eq.~(\ref{#1})}

\def\rcite#1{ref.~\cite{#1}}
\def\rcites#1{refs.~\cite{#1}}

\def\cA{{\cal A}}

\def\cL{{\cal L}}
\def\cM{{\cal M}}
\def\cN{{\cal N}}

\def\det{{\rm det}}

\def\å{\mathring{a}}

\newcommand\nn{\nonumber}

\newcommand{\Gmat}{\gamma}
\newcommand{\auxu}{z}
\newcommand{\auxv}{\tilde z}

\newcommand\spaq[1]{\langle #1\rangle}
\newcommand\spbq[1]{[#1]}

\def\spa#1.#2{\left\langle#1\,#2\right\rangle}
\def\spb#1.#2{\left[#1\,#2\right]}
\def\spab#1.#2{\left\langle#1\,#2\right]}
\def\spba#1.#2{\left[#1\,#2\right\rangle}
\def\spaq#1.#2{\langle#1 | #2\rangle}
\def\spbq#1.#2{[#1 | #2]}
\def\eqn#1{eq.~(\ref{#1})}

\def\bep{\boldsymbol{\mathbf{\ep}}}
\def\bep{{\varepsilon}}

\def\aket#1{\underline{\vphantom{q_3}\bm{#1}}\rangle}
\def\sket#1{\underline{\vphantom{q_3}\bm{#1}}]}
\def\ket#1{\underline{\vphantom{q_3}{#1}}\rangle}
\def\abra#1{\langle\underline{\vphantom{q_3}\bm{#1}}}
\def\sbra#1{[\underline{\vphantom{q_3}\bm{#1}}}

\def\gsug{{\rm g}}
\def\cubpol{{\cal V}}

\title{\boldmath Non-compact gauge groups, tensor fields and Yang-Mills-Einstein amplitudes}

\author[a]{Marco Chiodaroli,}
\author[b]{Murat G\"{u}naydin,}
\author[a,c]{Henrik Johansson,}
\author[b]{and Radu Roiban}

\affiliation[a]{Department of Physics and Astronomy,  \\ Uppsala University, Box 516, 75120 Uppsala, Sweden}
\affiliation[b]{Institute for Gravitation and the Cosmos, \\ The Pennsylvania State University,
University Park, PA 16802, USA}
\affiliation[c]{Nordita, Stockholm University and KTH Royal Institute of Technology,\\  Hannes Alfv\'{e}ns  v\"{a}g 12, 10691 Stockholm, Sweden}

\emailAdd{marco.chiodaroli@physics.uu.se}
\emailAdd{mgunaydin@psu.edu}
\emailAdd{henrik.johansson@physics.uu.se}
\emailAdd{radu@phys.psu.edu}

\abstract{Scattering amplitudes in Yang-Mills-Einstein theories have been investigated mostly for compact gauge groups.
While non-compact gauge groups are not physically viable in Yang-Mills theory, non-compact gaugings feature prominently in the supergravity literature, where any choice of perturbative vacuum spontaneously breaks the gauge group to a compact subgroup.
In this paper, we formulate double-copy constructions for several five-dimensional ${\cal N}=2$ supergravities with non-compact gauge groups. 
On one side of the double copy, we employ amplitudes from a super-Yang-Mills theory with a massive hypermultiplet. On the other, we use amplitudes from  particular non-supersymmetric Yang-Mills-scalar theories with massive fermions, chosen to obey constraints coming from color/kinematics duality.
Supergravities with massive self-dual tensors in five dimensions are also considered, showing that tensors are straightforwardly realized as double copies of gauge-theory fermions with suitable choices of signs in the corresponding solutions of the Dirac equation.  
We present several examples of these constructions, noting in particular the appearance of Heisenberg groups in the supergravity gauge symmetry and, in some cases, the possibility of exotic tensor-vector matter couplings. 
}

\preprint{UUITP–41/23 \\
\phantom{~} \hfill NORDITA-2023-081}

\begin{document}

\renewcommand{\thefootnote}{\arabic{footnote}}

\maketitle

\section{Introduction}

{Supersymmetric and non-supersymmetric Yang-Mills-Einstein theories (YME) combine gravitational and non-abelian gauge interactions and, as such, 
exhibit more structure than theories where all vector fields are abelian.}
YME scattering amplitudes  have been subject to intense investigation over the past decade, particularly in regards to their  relation with simpler amplitudes from Yang-Mills (YM) theory.
Building on earlier work~\cite{Bern:1999bx}, a more complete  understanding of these theories from the point of view of the so-called double-copy construction~\cite{Bern:2008qj,Bern:2010ue}  was first obtained by the current authors in \rcite{Chiodaroli:2014xia}. Subsequent work investigated amplitudes in YME theories building on the scattering equation formalism~\cite{Cachazo:2014nsa,Cachazo:2014xea},  string theory~\cite{Stieberger:2016lng}, and ambitwistor strings \cite{Casali:2015vta, Roehrig:2017wvh,Mazloumi:2022lga}.

YME amplitudes exhibit unexpected structure: they may be expressed as a weighted sum over amplitudes in YM theories, where the coefficients depend on the kinematic data ~\cite{Stieberger:2016lng, Nandan:2016pya,Schlotterer:2016cxa, delaCruz:2016gnm, Fu:2017uzt}.   Further investigation revealed that such amplitude relations can be understood as a consequence of an underlying double-copy structure \cite{Chiodaroli:2017ngp}, and led
to a general expression for $m$-point amplitudes~\cite{Teng:2017tbo,Du:2017gnh} (see also ref. \cite{Feng:2020jck}).
YME theories have a prominent place in the web of double-copy-constructible theories (see~\rcites{Bern:2019prr,Travaglini:2022uwo,Bern:2022wqg} for reviews) being also connected  with the Dirac-Born-Infeld and special Galileon theories by certain differential operators referred to as  transmutation operators~\cite{Cheung:2017ems, Feng:2019tvb}. For other work related to double-copy constructions that includes YME amplitudes, see refs.~\cite{Johansson:2017srf,Azevedo:2018dgo, Johansson:2018ues,Edison:2020ehu, Chen:2022nei, Xie:2022nfu}. For loop-level  results, see refs.~\cite{He:2016mzd,Nandan:2018ody,Faller:2018vdz, Edison:2020uzf, Porkert:2022efy}.

The ubiquity of the double copy among gravitational theories inspired the
suggestion \cite{Chiodaroli:2017ngp, Anastasiou:2017nsz} that perhaps all such theories have a double-copy property. 
The main focus  of the investigations leading to this conjecture has been   Maxwell-Einstein and YME theories with compact gauge groups.  
An analysis of gravitational theories that depart from this pattern has therefore the potential to further probe the breadth and validity of this proposal.

Supergravity theories possess non-compact duality groups (such as the $E_{7(7)}$ symmetry of maximal supergravity in four dimensions). An interesting property of these theories is that a non-compact subgroup of the duality group can be gauged without introducing ghost states in the spectrum. This is a genuinely novel feature of gravitational theories, which is not present in conventional gauge theories.
Another important characteristic of these theories is that they include fields transforming in nontrivial matter representations with respect to the gauge symmetry. 
Extensive studies led to a comprehensive understanding of YME theories in various dimensions from the Lagrangian perspective. 
In particular, five-dimensional Maxwell-Einstein theories with $\cN=2$ supersymmetry were investigated in~\rcites{Gunaydin:1983bi,Gunaydin:1984ak,Gunaydin:1986fg}, while YME and gauged supergravities were studied in~\rcites{Gunaydin:1984pf,Gunaydin:1985cu,Gunaydin:1999zx,Gunaydin:2000xk,Gunaydin:2003yx,Gunaydin:2005bf}. 
Despite their new physical features, YME theories are still much simpler than gauged supergravities, {\it i.e.} theories for which part of the R-symmetry is promoted to a local symmetry. For example, the scalar potential in YME theories either vanishes identically or at the critical point, guaranteeing the presence of Minkowski vacua.
Scattering-amplitude methods offer a new perspective on the rich structure of these theories and may reveal their place and connections in the web of double-copy-constructible theories.

In this paper, we initiate the study of YME theories with non-compact gauge groups from an amplitude perspective, focusing in particular on the double-copy approach. We draw from earlier results, which construct simpler YME theories as double copies in which the first gauge-theory factor is pure YM or super-Yang-Mills (SYM) theory and the second is a deformation of a non-supersymmetric  YM theory by trilinear scalar couplings (the so-called YM+$\phi^3$ theory) \cite{Chiodaroli:2014xia,Chiodaroli:2015rdg,Chiodaroli:2016jqw}. Focusing on $\cN=2$ supersymmetry and five spacetime dimensions, we show that this construction can be extended by including extra massive hypermultiplets in the supersymmetric gauge theory and extra fermions in the non-supersymmetric one, so that massive vectors corresponding to the non-compact generators of the gauge group and 
massive tensor fields are obtained as the double copies of two massive fermions. 
Two main additional tools are used to carry out these constructions. The first is the analysis of general constraints coming from color/kinematics (C/K) duality in the context of theories with massive fermions, which was first carried out with the intent to formulate double-copy constructions for gauged supergravities \cite{Chiodaroli:2018dbu}. The second is the novel extension of the spinor-helicity formalism to amplitudes in five spacetime dimensions \cite{Chiodaroli:2022ssi} (see also \rcites{Dennen:2009vk,Cheung:2009dc,Czech:2011dk,Cachazo:2018hqa,Geyer:2018xgb,Albonico:2020mge,Geyer:2020iwz}  for earlier closely related work).

Apart from being of interest in its own right, the study of YME theories with non-compact gauge groups from the double-copy perspective is also a stepping stone towards the analysis of the more challenging case of gauged supergravities  (theories in which R-symmetry is also gauged).
The double-copy construction for these theories was 
initiated in
\rcites{Chiodaroli:2017ehv,Chiodaroli:2018dbu}; the close connection between these two classes of theories is further underlined by the fact that they share a common gauge-theory factor in their construction. Amplitude methods have been successfully applied to many gravity and supergravity theories, showing great promise  \cite{Carrasco:2012ca,Chiodaroli:2014xia,Chiodaroli:2015rdg,Chiodaroli:2015wal,Chiodaroli:2016jqw,Anastasiou:2016csv,Chiodaroli:2017ngp,Ben-Shahar:2018uie,Chiodaroli:2017ehv,Anastasiou:2017nsz} (see also \rcite{Bern:2019prr} and \rcites{Travaglini:2022uwo,Bern:2022wqg,Adamo:2022dcm} for  reviews).
Both in case of YME theories and gauged supergravities, it is therefore hoped that the double-copy approach can help identify and study new families of theories ({\it i.e.} new gaugings), connecting amplitudes methods with very recent results from the supergravity community \cite{Dallagata:2021lsc,Bobev:2020ttg,Krishnan:2020sfg}. 

In this paper, we provide double-copy constructions for several gaugings of the generic Jordan family, the complex magical supergravity theory, and the quaternionic magical supergravity theory. Most of these constructions are based on the introduction of massive fermions in the non-supersymmetric gauge theory in a way that preserves C/K duality. Table \ref{tabgaugings} summarizes the main gaugings discussed here.

\begin{table}

\begin{tabular}{c|c@{}c@{}c}
\bf Theory & & \bf Gaugings & \\[3pt]
\hline
Generic Jordan  & $SO(2,1)$ & $U(1)\ltimes {\cal H}_3$  & $U(1)$ gauging \\
family          & gauging   & gauging                   & with tensors   \\[3pt]
\hline
Complex   & $U(2)\ltimes {\cal H}_5$ & $U(2)$ gauging  & $U(1,1)\ltimes {\cal H}_3$ gauging \\
magical sugra    & gauging   & with tensors   & with tensors   \\[3pt]
\hline
Quaternionic  & $U(2)\ltimes {\cal H}_5$ gauging & \ $U(1,1)\ltimes {\cal H}_5$ gauging  &  \\
magical sugra  & gauging with tensors  & with tensors   &  \\

\end{tabular}
\caption{List of gaugings constructed as double copies in this paper.\label{tabgaugings}}
\end{table}

This paper is structured as follows. In Section \ref{sec2}, we give a quick review on $\cN=2$ YME theories in five dimensions, summarizing the relevant results for the current investigation and referring the reader to the original literature for details. In Section~\ref{sec3}, we give the generalities of the double-copy construction for non-compact YME theories. We discuss our main technical tools, such as C/K duality in theories with massive fermions and the choice of gauge-group representations for the theories entering the double copy. We show that different pairings between gauge-theory states give rise either to massive vectors or to massive tensors, the latter possibility only existing in five or higher dimensions. We moreover include a summary of the main  constructions presented in this paper.
Sections~\ref{sec3}-\ref{sec5} contain details about the above constructions, including matching with amplitudes from supergravity Lagrangians. We focus separately on  the generic Jordan family, the complex magical theory and the quaternionic magical theory. We conclude 
with a discussion of directions for future work, while the appendices contain a summary of the five-dimensional spinor-helicity formalism we employ throughout the paper and details on the choice of Gamma matrices.

\section{Review of non-compact Yang-Mills-Einstein theories \label{sec2}}

We start by reviewing some fundamental results on Maxwell-Einstein and YME theories with $\cN=2$ supersymmetry in five dimensions. This section is meant as a non-comprehensive summary of results that will be used in later sections to relate the output of the double copy with amplitudes from the supergravity Lagrangian. We refer the reader to the original literature, and in particular to~\rcites{Gunaydin:1983bi,Gunaydin:1984nt,Gunaydin:1984pf, Gunaydin:1984ak,Gunaydin:1999zx, Gunaydin:1986fg}, for an exhaustive treatment.

\subsection{General five-dimensional Maxwell-Einstein Lagrangian \label{sec5dMESGT}}

Our starting point is a five-dimensional theory involving the $\cN=2$ graviton multiplet and $n_V$ vector multiplets. The bosonic part of the $\cN=2$ Maxwell-Einstein supergravity Lagrangian is the following,\footnote{Supergravity Lagrangians are given with mostly-plus metric signature throughout the paper, following a standard choice in the supergravity literature.}
\bea
e^{-1} {\cal L} &=& -{1 \over 2}R -{1\over 4} \å_{IJ} F^I_{\mu \nu} F^{J \mu \nu }   
- {1 \over 2} g_{xy} \partial_\mu \varphi^x \partial^\mu \varphi^y+{e^{-1} \over 6 \sqrt{6} } C_{IJK} \epsilon^{\mu \nu \rho \sigma \lambda} F^{I}_{\mu \nu}
F^J_{\rho \sigma} A^K_{\lambda} \ , \label{ungaugedL} \qquad \eea
where $A^I$ with  $I=0,\ldots,n_V$ are the vectors in the theory and $\varphi^x$, $x=1,\ldots n_V$ are real scalars, which can be seen as the coordinates of the scalar manifold  ${\cal M}_{\rm 5D}$ with metric  $g_{xy}$.
$C_{IJK}$ are symmetric constant tensors.
{A key result in the supergravity literature is that the $C_{IJK}$ tensor specifies completely the $\cN=2$ Maxwell-Einstein supergravity }theory, {\it i.e.} all other objects, including the ones appearing in the fermionic part of the Lagrangian, are expressible in terms of this tensor \cite{Gunaydin:1983bi}. This is done as follows. We first use the $C_{IJK}$ tensor to write an associated cubic polynomial in some auxiliary coordinates $\xi^I$, 
\be
{\cubpol}(\xi) = \Big( {2 \over 3}\Big)^{3/2} C_{IJK} \xi^I \xi^J \xi^K \ . 
\ee
We then define the scalar manifold ${\cal M}_{\rm 5D}$  by the equation $\cubpol(\xi^I)=1$ in the ambient space spanned by the $\xi^I$ coordinates. This ambient space can be naturally endowed with the metric
\be 
a_{IJ} = -{1 \over 2} \partial_I \partial_J \ln \cubpol (\xi) \  .
\ee
The metric in the kinetic-energy  term for the vector fields is then  the restriction of $a_{IJ}$ 
to the 
scalar manifold ${\cal M}_{\rm 5D}$, while
the scalar-manifold metric $g_{xy}$ is the induced metric on ${\cal M}_{\rm 5D}$,
\bea 
\å_{IJ}(\varphi) = a_{IJ} \big|_{{\cubpol}(\xi)=1} 
\ , \qquad  \qquad 
g_{xy}(\varphi) = \å_{IJ} \partial_x \xi^I \partial_y \xi^J \ . 
\eea
{All the above constraints follow from the local supersymmetry of the Lagrangian. }

It will be useful to express $\å_{IJ}$ in terms of the appropriate vielbeine and their inverses, which obey the relations
\bea 
\å_{IJ}= h_I h_J + h^a_I h^a_J \ , & \qquad & \no 
h^I_a  h^J_b \å_{IJ} = \delta_{ab} \ , \\  
h^I h^J \å_{IJ} = 1  \ , & \qquad &
h^I h_{Ia} = h_I h^{Ia}=0 \ .
\eea
In turn, these objects are readily expressed in terms of the ambient-space coordinates,
\bea 
h^I(\varphi) = \sqrt{ 2 \over 3} \xi^I \big|_{{\cubpol}(\xi)=1}  &\qquad&
h_I(\varphi) = \sqrt{ 1 \over 6} \partial_I \ln {\cubpol}(\xi) \big|_{{\cubpol}(\xi)=1} \ ,\nn  \\
h_{x}^I(\varphi) = - \sqrt{3 \over 2} \partial_x h^I \ , & \qquad& 
h_{Ix} (\varphi) = \sqrt{3 \over 2} \partial_x h_I 
\ .
\eea
The $C_{IJK}$ tensors are  not completely free. Indeed, they are constrained 
by the requirement that the kinetic terms for the physical fields must  be positive-definite. 
To carry out amplitude calculations it is necessary to expand the action around a base-point $c^I$ of the scalar manifold ${\cal M}_{\rm 5D}$, {\it i.e.} an expectation value for the scalar fields, $\xi^I=c^I+\ldots$, compatible with $\cubpol(c^I)=1$. 
Choosing the base-point as 
\be 
c^I = \Big( \sqrt{3 \over 2}, 0, \ldots, 0 \Big)  \ ,
\ee
one finds that the most general $C_{IJK}$ tensor compatible with positive kinetic terms can be brought to the form
 \be 
 C_{000} = 1 , \qquad  C_{0ab}= -{1 \over 2} \delta_{ab} , \qquad C_{00a} = 0,  \qquad  a,b,c=1,\ldots,  n_V  \ ,  \label{Ccan} 
 \ee
with the remaining entries {$C_{abc}$ of the $C_{IJK}$ tensor being completely arbitrary.}

From this discussion, it is clear that the invariance group of the   $C_{IJK}$ tensor becomes the global symmetry group of the whole theory. { In this paper, we will focus mainly on theories which possess symmetric scalar manifolds.}  The simplest group of theories of this class is the so-called generic Jordan family, with scalar manifold in five dimensions
 \be {\cal M}_{\rm 5D} = {SO( n_V-1,1)\times SO(1,1) \over SO( n_V-1)} \ , \qquad  n_V \geq 1 , \ee
where again $n_V$ is the number of vector multiplets. Gaugings for these theories will be studied in Section \ref{sec4}. { For completeness, we also mention the generic non-Jordan family of theories, which has scalar manifolds \cite{Gunaydin:1984pf} 
\be 
\mathcal{M}_{\rm 5D} = {SO(n_V,1)\over SO(n_V)} \ .
\ee 
In this case, interactions break the isometry group down to the parabolic subgroup  
\be \big[
SO(n_V-1)\times SO(1,1)\big] \ltimes \mathbb{T}^{n_V{-}1} \ .
\ee
In addition to these two families, we have four isolated cases known as the Magical supergravities which are  defined by the degree-three simple Euclidean Jordan algebras over the four division algebras \cite{Gunaydin:1983bi}. They have the scalar manifolds
\bea 
\mathcal{M}_{\rm 5D}^\mathbb{R}= \frac{SL(3,\mathbb{R})}{ SO(3)}, & \qquad &  \mathcal{M}_{\rm 5D}^{\mathbb{H}}=\frac{SU^*(6)}{USp(6)},\nn \\ 
\mathcal{M}_{\rm 5D}^{\mathbb{C}} =\frac{SL(3,\mathbb{C})}{ SU(3)},  & \qquad &  \mathcal{M}_{\rm 5D}^{\mathbb{O}}= \frac{E_{6(-26)}}{F_4} \ , 
\eea
and describe the coupling of 5, 8, 14 and 26 vector multiplets to $\cN=2$ supergravity in five dimensions, respectively. Some of their gaugings will be studied in Sections~\ref{sec5}-\ref{sec6}. We should also note that complex, quaternionic and octonionic magical supergravities belong to three infinite families of unified Maxwell-Einstein supergravity theories~\cite{Gunaydin:2003yx}. The scalar manifolds of the other theories in these unified families are neither symmetric nor homogeneous, and will not be studied in this paper. 

A family of supergravities with homogeneous target spaces, in which the symmetric space theories are a particular subset, was classified in \rcite{deWit:1991nm}. These theories form an infinite family identified by two integers $q\geq -1 $ and $P\geq 0$, with the option of a third parameter $\dot P \geq 0$ appearing for $q= 0$ (mod $4$). The corresponding isometry groups are
\begin{equation}
\big[SO(1, 1) \times SO(q + 1, 1) \times {\cal S}_q(P, \dot P) \big]
\ltimes \mathbb{T}^{(P+ \dot P)D_q} \ ,
\end{equation}
where ${\cal S}_q(P, \dot P) $ is a flavor group depending on the value of $q$ and $D_q$ gives the number of  five-dimensional Dirac spinors corresponding to an irreducible spinor in $(q+6)$ dimensions. The corresponding scalar manifolds are homogeneous spaces and have been dubbed $L(q,P)$ and $L(q,P, \dot P)$ spaces in the supergravity literature  \cite{deWit:1991nm}.

\subsection{Yang-Mills-Einstein theories with and without tensors \label{LagrangianDiscussion}}

YME theories are obtained by promoting a subgroup of the  global symmetry group of a Maxwell-Einstein theory to a local gauge symmetry. These theories are often times referred to as ``gaugings" of the original Maxwell-Einstein supergravity.
More concretely, the global invariance group $G$ of a $\cN=2$ Maxwell-Einstein supergravity is the group of transformations leaving the  $C_{IJK}$ tensor invariant. We then identify a subgroup $K$ of  $G$ which acts infinitesimally in the ambient space as 
\be \delta_{\alpha} \xi^I = (T_s)^I_{\ J} \xi^J \alpha^r \ ,  \qquad 
(T_s)_{(I}^{~L} C_{JK)L} =0 \ , \label{transformations}
\ee
where $T_s$ are the generators to be gauged and $\alpha^s$ are the transformation parameters. The scalars and vectors transform as 
\bea \delta_\alpha \varphi^x = K^x_s \alpha^s \ , & \qquad &
\delta_{\alpha} A^I_\mu  =  (T_s)^I_{\ J} A^J_\mu \alpha^s \ ,
\eea
where  $K^x_s$ is the Killing vector corresponding to the action of the $s$-th generator on the scalar manifold, 
\bea  K^x_s &=& - \sqrt{ 3 \over 2} (T_s)^J_{\ I} h_J h^{I x} .  \label{defK}
\eea
We will not need the transformations on the fermionic fields, for which the reader is referred to the original literature  \cite{Gunaydin:1983bi}. 

To promote $K$ to a local symmetry, we need the adjoint representation of $K$ to be embedded in the  representation of the global symmetry $G$ under which the vector fields transform. We label the corresponding vectors as $A^s_\mu$, with the  index $s$ running over the adjoint representation of the gauge group $K$  plus additional singlets. In the general case, there will be vectors transforming in additional nontrivial representations of the gauge group. One of the fundamental results from the supergravity literature \cite{Gunaydin:1999zx} is that, in order for the theory to preserve local $\cN=2$ supersymmetry, the field strengths of these additional vectors must be 
dualized to massive tensor fields, which will be denoted as $B_{\mu \nu}^N$, where the original index $I$ is split as $I=(s,N)$, $J=(t,P)$ and these massive tensors satisfy odd-dimensional self-duality conditions~\cite{Gunaydin:1999zx}.

In general, the representation matrices $T_s$ will be written as 
\begin{equation}
(T_s)^I_{\ J} = \left( \begin{array}{cc} 
f^t_{\ s u} & 0 \\ 0 & (T_{s})^N_{\ P} 
\end{array}\right) \ , \label{Mmatrix}
\end{equation}
where $(T_s)^N_{\ P}$ will become representation matrices for the massive tensors under the gauge group. These are obtained from the entries of the $C_{IJK}$ tensor in terms of an antisymmetric   matrix $\Omega^{NP}$, which is interpreted as a symplectic metric ($\Omega^{NP}\Omega_{PQ}=\delta^N_Q$), as 
\begin{equation}
 (T_s)^N_{\ P} = \sqrt{2 \over 3} \Omega^{NQ} C_{sQP} \ .
\end{equation}
By construction, the representation matrices are symplectic with respect to  the inverse symplectic metric $\Omega_{NP}$,
\begin{equation}
(T_s)^Q_{\ N} \Omega_{QP} + (T_s)^Q_{\ P} \Omega_{NQ} = 0 \ .
\end{equation}
We can then group vector field strengths and tensors together defining 
\begin{equation}
{\cal H}_{\mu \nu}^I = ({\cal F}^s_{\mu \nu} , B^N_{\mu \nu}) \ ,
\end{equation}
where ${\cal F}^s_{\mu \nu}= F^s_{\mu \nu} + \gsug f^s_{\ tu} A^t_\mu A^u_\nu$ are non-abelian field strengths and $\gsug$ denotes the coupling constant for supergravity gauge group.\footnote{The supergravity coupling constant $\gsug$ was denoted as $g_s$ in ref. \cite{Chiodaroli:2022ssi}.} The Lagrangian of the desired YME theory with tensors is then obtained from eq. (\ref{ungaugedL}) by covariantizing derivatives and suitably extending the $F\wedge F \wedge A$ term;
the bosonic sector  has the Lagrangian~\cite{Gunaydin:1999zx}
\bea
 \label{gaugedL} 
e^{-1} {\cal L} &=& -{R \over 2} -{1\over 4} \å_{IJ} \mathcal{H}^I_{\mu \nu} \mathcal{H}^{J \mu \nu }   
- {1 \over 2} g_{xy} \mathcal{D}_\mu \varphi^x \mathcal{D}^\mu \varphi^y +  \no 
{e^{-1} \over 6\sqrt{6}} C_{stu} \epsilon^{\mu\nu\rho\sigma\lambda} \left\{ \vphantom{1\over 2}
 F^s_{\mu \nu} F^t_{\rho \sigma} A^u_{\lambda} \right. \\
&& \quad \left. + {3 \over 2} \gsug F^s_{\mu \nu} f^{t}{}_{ t' u'}  A^u_\rho A^{t'}_\sigma A^{u'}_\lambda +
{3\over 5} \gsug^2 A^s_\mu f^{t}{}_{ s' t'} A^{s'}_\nu A^{t'}_\rho f^{u}{}_{u' v'} A^{u'}_\sigma A^{v'}_{\lambda} \right\}   \no\\
&& \quad + {e^{-1} \over 4 \gsug} \epsilon^{\mu\nu\rho\sigma\lambda} \ \Omega_{NP} B_{\mu \nu}^N {\cal D}_\rho B_{\sigma \lambda}^P - {9 \over 8} \gsug^2 \å_{SN} (T_s)^S_{\ P} h^s h^P \, (T_t)^N_{\ Q} h^t h^Q \ ,    \eea
where the covariant derivatives for scalars and tensors are 
\begin{eqnarray}
{\cal D}_\mu \varphi^x &=& \partial_\mu \varphi^x + \gsug A^s_\mu K^x_s \ , \nn \\
{\cal D}_{[\mu} B^N_{\nu \rho]} &=& \partial_{[\mu} B^N_{\nu \rho]} + \gsug  (T_s)^N_{\ P} A^s_{[\mu} B^P_{\nu \rho]} \ \ .
\end{eqnarray}
In this paper, we will consider both cases in which $K$ is compact and cases in which it is non-compact.

Finally, the block diagonal structure in eq. (\ref{Mmatrix}) can be further generalized to include an off-diagonal block,
\begin{equation}
(T_s)^I_{\ J} = \left( \begin{array}{cc} 
f^t_{\ s u} & (T_{s})^t_{\ P} \\ 0 & (T_{s})^N_{\ P} 
\end{array}\right) \ . \label{Mmatrixskrewy}
\end{equation}
In \rcite{Gunaydin:1999zx}, it was pointed out that such a generalization appears to be impossible to supersymmetrize in a gauge-invariant way except possibly in very special cases. 
These correspond to working with reducible but not completely reducible representations and producing  new vector-tensor couplings 
which were studied in \rcite{Bergshoeff:2004kh}. For compact gaugings, such an extension is not possible since all their reducible unitary representations are fully reducible. Similarly, every reducible representation of a  connected semisimple non-compact Lie group is also fully reducible \cite{Cornwell:1997ke}. Hence, the only way to extend gaugings to reducible representations of the form (\ref{Mmatrixskrewy}) is to gauge a non-compact Lie algebra with an abelian invariant subalgebra~\cite{Bergshoeff:2004kh}.  We shall see that such exotic gaugings appear naturally in the framework of the double-copy construction.

\subsection{Matching amplitudes and spectra with the supergravity Lagrangian \label{ampgrav}}

The first step to find double-copy realizations for the theories described in the previous subsections is to identify parameters to match between the output of the double copy and the Lagrangian description of the supergravities under consideration. In general, to compute amplitudes directly, we will need to expand the Lagrangian around a base-point in the scalar manifold, which we denote as $\xi^I=c^I$ in ambient space.
Inspecting the terms of the Lagrangian proportional to the coupling constant $\gsug$ we then have
\begin{eqnarray}
- \gsug \, \å_{st} \big|_0  \partial_{[\mu}A^s_{\nu]} f^t_{\ uv} A^{u\mu} A^{v\nu} - \gsug \, g_{xy} K^y_s \big|_0 D_\mu \varphi^x   A^{s \mu}   \ .
\end{eqnarray}
The latter term gives a non-diagonal contribution to the quadratic Lagrangian, which we cancel with the gauge-fixing term
\begin{equation}
{\cal L}_{\rm gf} = -{1 \over 2 \xi } \Big( D^\mu A^t_\mu + \xi \, \gsug \, g_{xy} K^{ty} \big|_0 \varphi^x \Big)^2 \ ,
\end{equation} 
where $\xi$, not to be confused with the ambient-space coordinates, is the parameter in the $R_\xi$ gauges, and $g_{xy}$, $K^{ty}$ are evaluated at the base-point. Using this procedure, we can write mass terms for the vector fields as 
\begin{equation}
M^2_{st} = \gsug^2 g_{xy} K^x_s K^y_t \big|_0 \ ,
\end{equation}
where we have used the definition (\ref{defK}).\footnote{{We should note that this mass term is non-zero only for the non-compact gauge fields.} } The masses of the tensor fields, on the other hand, are given by the entries of the $\å$ matrix evaluated at the base-point, 
\begin{equation}
M^2_{NP} = \å_{NP} \big|_0 \ .
\end{equation}
While a comparison of the mass spectrum is the first nontrivial check that a proposed double-copy construction must pass, additional important information is in the three-point amplitudes. In particular, three-point amplitudes between one vector and two tensors are of the form\footnote{To have all three-point amplitudes exhibit the same mass dimension, we redefine the tensor fields with respect to the supergravity Lagrangian so that an explicit factor of $m^2$ appears in the amplitude.}
\begin{equation}
{\cal M}_3(1 B^N,2 B^P,3 A^s,) =   i \sqrt{8 \over 3} \å^{st}_0 \Omega^{NR} \Omega^{PS} C_{tRS} \, m^2 \epsilon \big( \varepsilon^B_1, \varepsilon^B_2 , \varepsilon_3 \big) \ ,
\label{BBAamp}
\end{equation}
where the Levi-Civita contraction between a vector and two tensors is defined in the natural way $\epsilon(a, b, c)\equiv \epsilon^{\mu\nu\rho\sigma\zeta}a_{\mu} b_{\nu\rho} c_{\sigma\zeta}$, 
and $\å^{st}_0 \equiv \å^{st}\big|_0$ are inverses of the matrix in the kinetic term for the vectors evaluated at the base-point (for practical calculations it is often convenient to redefine fields so that their kinetic terms are 
canonically normalized).
Amplitudes between three vectors have the form\footnote{We note that this amplitude is not modified when the metric signature is changed to mostly minus.}
\bea
{\cal M}_3(1A^s,2A^t,3A^u)   &=&  i  { \sqrt{ 8 \over 3}} \å^{ss'}_0 \å^{tt'}_0 \å^{uu'}_0  C_{s't'u'} \epsilon \big(\varepsilon_1, \varepsilon_2, \varepsilon_3, p_1, p_2 \big)   \\
&& \hskip -1cm \null  + \gsug \big\{ \å^{tt'}_0\å^{uu'}_0  f^s_{\ t'u'} ( \varepsilon_1 \cdot  \varepsilon_2 \, p_1 \cdot  \varepsilon_3 - \varepsilon_1 \cdot  \varepsilon_3 \, p_1 \cdot  \varepsilon_2) + \text{cyclic}(1,2,3) \big\} \ . \no
\label{3ptsugravec}
\eea
The first term is the  contribution of the $F\wedge F \wedge A$ term which is quadratic in the momenta and non-vanishing if the gauge coupling constant is set to zero. We have
$\epsilon(a,b,c,d,f) \equiv \epsilon^{\mu\nu \rho \sigma \lambda} a_{\mu}b_{\nu}c_{\rho}d_{\sigma}f_{\lambda}$. Note that this contribution is always symmetric in the three indices.
The second contribution originates from the covariantization of the derivatives in the YME Lagrangian.  
Unless the $\å^{IJ}$ already happens to be canonically normalized at the base-point and the gauge group is compact, the color tensors will not be antisymmetric in all three indices. This is an important difference with respect to the gauge-theory case. To inspect three-vector amplitudes more in detail, we focus on the case in which the third  vector is massless, while the first two are massive.\footnote{In analogy with  dimensional compactification, in this paper, we take masses to carry a sign so that amplitudes are supported by asymptotic states obeying the condition $\sum_i m_i=0$. Our approach follows the one from ref. \cite{Chiodaroli:2022ssi}, but some signs are different in this paper as we choose conventions in which all momenta of the asymptotic states are taken to be incoming.} In this case, two different gauge-invariant terms with linear dependence on momenta appear in the Lagrangian. The first amplitude we need to consider originates from the operator
\begin{equation}
{\cal O}_1=  - 2 {D_{[\mu} \overline{W}_{\nu]}} D^\mu W^\nu \ .
\end{equation}
This leads to a contribution to the three-point amplitude which is proportional to 
\bea
 \langle \overline W  W A \rangle \Big|_{|D W|^2} ~&=&~i ( \bep_1 \cdot  \bep_2 \, (p_1-p_2) \cdot  \varepsilon_3 +  \bep_2  \cdot \varepsilon_3 \, p_2\cdot  \bep_1 -\bep_1  \cdot \varepsilon_3 \, p_1\cdot \bep_2 )  \\
&&  \hskip -3cm
=  {i \over 2m^2} \left\{ \big(    \abra{1}\aket{2} \, \sbra{2}\sket{1} - \abra{2}\sket{1} \, \abra{1}\sket{2} \, \big) (p_1 \cdot \varepsilon_3) -
 {1 \over 4} \Big(\abra{1} | p_2 |\sket{1} \, \abra{2} | \varepsilon_3 | \sket{2}  - (1 \leftrightarrow 2)  \Big) \right\} , \nn \label{O1expanded}
\eea
where we have employed the five-dimensional spinor-helicity formalism in the last line, and we have used the short-hand notation $\abra{1} | p_2 |\sket{1} \equiv \abra{1} | \cancel{p}_2 |\sket{1}$ (with a similar notation  for other combinations of angle and square brackets). This will be used to simplify expressions throughout the paper. Relevant identities are summarized in Appendix \ref{AppA}, while the reader is referred to \rcite{Chiodaroli:2022ssi} for a complete treatment.
The second gauge-invariant term in the supergravity Lagrangian has the form 
\be
{\cal O}_2=  - i F^a_{\mu \nu}\overline{W}^\mu t^a W^\nu \ .
\ee
Its contribution is proportional to
\bea
  \langle \overline W W A \rangle \Big|_{\overline{W} \cdot F \cdot {W} } ~&=& i  (\bep_1 \cdot \varepsilon_3 \, p_3 \cdot \bep_2 -\bep_2 \cdot \varepsilon_3 \, p_3 \cdot  \bep_1  )  \nn \\
 &=&  -{i \over 8m^2}  \Big(\abra{1} | p_2 |\sket{1} \, \abra{2} | \varepsilon_3 | \sket{2}  - (1 \leftrightarrow 2)  \Big)   \ . \label{O2expanded}
\eea
These operators are sufficient to account for the interactions introduced in the supergravity theory by the gauging procedure. In addition to these, we should note that the ungauged theory already has a non-zero three-point amplitude.
The corresponding contribution to the two-massive-one-massless three-vector amplitude is
\be
  \langle  \overline W  W A \rangle \Big|_{ D\overline{W}  \wedge DW \wedge A} =
4 i \, \epsilon(\bep_1, \bep_2, \varepsilon_3, p_1, p_2) =
i  \Big(  \abra{1}\aket{2} \,  \sbra{1} | \varepsilon_3 | \sket{2} +  \sbra{1}\sket{2} \,  \abra{1} | \varepsilon_3 | \aket{2}  \Big)  ,
\ee
which is proportional to the $C_{IJK}$-tensor of the theory.

In a concrete calculation, massive and massless vectors will carry additional indices of the local (or sometimes global) symmetries of the supergravity theory under consideration. Hence, eqs.~(\ref{O1expanded}) and (\ref{O2expanded}) each give the kinematical part of a three-point amplitude which also includes a color tensor. If we impose  antisymmetry in the two massive vector fields, eqs.~(\ref{O1expanded}) and (\ref{O2expanded}) are the only independent three-point matrix elements  
that can appear. By explicit comparison with   (\ref{3ptsugravec}) we can see that ${\cal O}_1$ is associated to the structure constants $f^{s,tu} = - f^{t,su} $, while ${\cal O}_2$ carries a color factor of the form $f^{u,st}$; here we have used the matrix $\å^{IJ}_0$ at the base-point to raise two of the indices of the structure constants.   
Note that in the YM Lagrangian with spontaneously-broken gauge symmetry both terms are present with the same normalization. However, since 
the kinetic terms of the vector fields are not necessarily canonically normalized at the base-point, and the gauge groups we are considering are not necessarily semi-simple, these two terms can and will occur with different weights.

A simple and important  example of a setting in which only ${\cal O}_2$ appears is given by the case of the Heisenberg group ${\cal H}_{2P+1}$; in this case the algebra of the generators has the form
\begin{equation}
[ {\bf q}^i, {\bf p}^j ] = Z \delta^{ij} \ , \qquad [{\bf p}^i, {\bf p}^j] = [{\bf p}^i, Z] = [{\bf q}^i, {\bf q}^j] = [{\bf q}^i, Z] = 0 \ , \qquad  i,j= 1,\ldots, P \, ,
\end{equation}
where $Z$ is the generator associated to the central charge of the algebra. If ${\bf q}^i$ and ${\bf p}^i$ are Hermitian operators, the central charge $Z$ must be anti-Hermitian. It is immediate to read off the nonzero structure constants, 
\begin{equation}
f^Z{}_{{\bf q}^i , {\bf p}^j} = - f^Z{}_{{\bf p}^j,{\bf q}^i}  = \delta^{ij} \ .
\end{equation}     
In this case, provided that the matrix in the kinetic term for the vectors is canonically normalized at the base-point, only the operator ${\cal O}_2$ appears in the three-point amplitudes. As we shall see, the appearance of a Heisenberg group is a common feature of many of the gaugings obtained from the double-copy construction in this work.

\section{Double-copy construction: generalities \label{sec3}}

We are now ready to describe the main features of the double-copy constructions considered in this paper. The starting point is the construction for homogeneous Maxwell-Einstein theories discussed in \rcite{Chiodaroli:2015wal},
\begin{align}
\left(
\begin{array}{c}
\cN  =   2  \text{ homogeneous} \\
\text{ supergravity }
\end{array} \right)
\quad = \quad  
\left(
\begin{array}{c}
 \cN = 2 \text{ SYM}  \\
 \null + \frac{1}{2} \text{ hyper}_R
\end{array} \right) 
 \otimes
\left(
\begin{array}{c}
\text{YM}+ n_s \text{ scalars}  \\ \null + n_F \text{ fermions}_R
\end{array} \right)  \,,  \label{DChom}
\end{align}
{\it i.e.} a double-copy construction in which the first gauge theory is a supersymmetric theory with a matter half-hypermultiplet, and the second theory is a higher-dimensional YM theory with matter fermions reduced to five dimensions.

Since we are seeking to construct theories with various massive fields, it is clear that we need to consider massive deformations of these gauge theories. The deformation for the supersymmetric theory is straightforwardly obtained by adding a mass term for the half-hypermultiplet. There is considerably more freedom as far as the non-supersymmetric gauge theory is concerned. A very natural starting point is  a massive deformation  which also involves trilinear scalar couplings, for which the Lagrangian is  
\bea
{\cal L} &=& - {1 \over 4} (F^{\hat a}_{\mu \nu})^2 + {1 \over 2} (D_\mu \phi^{\hat a i})^2 
-{1\over 2 } m^2_{ij}  \phi^{\hat a i}  \phi^{\hat a j} \!\!
- {g^2 \over 4} f^{\hat a \hat b \hat e} f^{\hat c \hat d \hat e} \phi^{\hat a i} \phi^{\hat b j} \phi^{\hat c i} \phi^{\hat d j}    \\
&& \null + {i \over 2} \bar \psi \cancel{D} \psi - {1 \over 2}  \bar \psi M \psi  + {g \over 2} \phi^{\hat a i} \bar \psi \Gamma^{4+i} t_R^{\hat a} \psi - {g\lambda \over 3!} f^{\hat a \hat b \hat c} F^{4+i,4+j,4+k} \phi^{\hat a i}  \phi^{\hat b j}  \phi^{\hat c k} \, ; \nn
\label{actionGSGgt}
\eea
here $\hat a, \hat b, \hat c$ are gauge-group adjoint indices, $i,j,k=1,\ldots,n_s$ are global indices running over the scalars. Fermions $\psi$ transform in a representation with representation matrices $t_R^{\hat a}$. $\Gamma^I$ are matrices appearing in the Yukawa couplings 
and can be seen as gamma matrices for the YM theory in $D=n_s+5$ dimensions, as explained in Appendix \ref{appB}.  Gauge and global indices carried by the fermions are not explicitly displayed. This theory was studied in \rcite{Chiodaroli:2018dbu}, where it was shown that  
C/K duality of the two-scalar-two-fermion four-point amplitude requires that the fermion mass matrix $M$ obey the relation
\be 
\big[ \Gamma^I , \big\{ \Gamma^J , M  \big\} \big]  + i \lambda F^{IJK} \Gamma^K  = 0 \,, \label{masterintro} 
\ee
where $\Gamma^I$ are the Dirac matrices in higher dimensions and $F^{IJK}$ are related to the structure constants of the supergravity gauge group. This constraint can be solved by the following choice for the fermion masses,
\be
M= u_l \Gamma^l+i {\lambda \over 4! }  F_{ijk} \Gamma^{ijk} \,,
\label{fermionmass}
\ee
where $u^I$ are extra parameters and $\Gamma^{ijk}$ is the rank-three element of the Clifford algebra.
This theory can be further enriched by going on the Coulomb branch, which is necessary for exploring the moduli space of the corresponding supergravities. Finally, an orbifold projection can be used to eliminate some of the fields. These two latter steps will be enacted on a case-by-case basis throughout the following sections.

Before outlining the details of all the possible constructions, we need to first comment on two distinct ways to double copy two gauge-theory fermions.
Using the five-dimensional spinor-helicity formalism summarized in Appendix~\ref{AppA}, we can write the external spinor polarizations as $|{\bm p}^a \rangle$ and $|{\bm p}^{\dot a} ]$, where $a,{\dot a}$ are little group indices. The key observation is that the double copy gives different outputs according to the way different fermionic states from the gauge theories are paired.	
\begin{enumerate}
	\item One possibility is to associate supergravity states to fermion bilinears of the form
	\begin{equation}
	 |{\bm p}^a \rangle \otimes |{\bm p}^{\dot a} ]  \ ,
	 \qquad 
	 |{\bm p}^{\dot a} ] \otimes |{\bm p}^{ a} \rangle  
	\end{equation}
 {\it i.e.} the spinor polarization from the different sides of the double copy obey the Dirac equation with opposite sign for the mass. 
\item Another possibility is to associate supergravity states to fermion bilinears of the form
	\begin{equation}
	 |{\bm p}^a \rangle \otimes |{\bm p}^{b} \rangle  \ ,
	 \qquad 
	 |{\bm p}^{\dot a} ] \otimes |{\bm p}^{\dot a} ] 
	\end{equation}
	{\it i.e.} the spinor polarization from the different sides of the double copy obey the Dirac equation with same sign for the mass. 
\end{enumerate}
In five dimensions, as the little-group indices suggest, the above choices leads to different fields in the double copy. The first leads to a five-dimensional vector, with polarization
\be
\varepsilon^\mu_{a  \dot a} (p) =  -\frac{ \langle {\bm p}_{a}| \Gmat^\mu   | {\bm p}_{{\dot a}}]}{2\sqrt{2}m}\, .
\label{massivepol}
\ee
The second choice leads to a five-dimensional massive tensor field plus a scalar. The tensor is given by the symmetric combination of little-group indices, and their 
polarization are
\begin{align}
\varepsilon^{\mu\nu}_{a  b} (p) =  \frac{ \langle {\bm p}_{a}| \Gmat^{\mu \nu }  | {\bm p}_{b} \rangle}{4\sqrt{2}m}\ , 
\qquad \qquad
\varepsilon^{\mu\nu}_{\dot a  \dot b} (p) =  \frac{ [{\bm p}_{\dot a}| \Gmat^{\mu \nu }  | {\bm p}_{\dot b} ]}{4\sqrt{2}m}\, .
\label{poltensors}
\end{align}
The additional state, given by the antisymmetric combination of little-group indices, is a scalar, which bringing the total number of degrees of freedom to four. In the presence of $\cN=2$ supersymmetry, these tensor and scalar fields belong to a massive tensor multiplets. On the other side, the massive vector field belongs to a massive vector multiplet which is BPS and does not have any scalars.  The properties of these polarization tensors and vectors are summarized in Appendix \ref{AppA}.

In practical situations, the gauge-group representations of the fermions in the two gauge theories dictate which states should be combined under the double copy, according to the criterion that a supergravity state is associated to a gauge-invariant bilinear of gauge-theory states.  

In contrast to the construction in \rcite{Chiodaroli:2015rdg} for massless theories, when introducing masses, it will be more convenient to consider complex gauge-group representations.  The reason for this choice can be traced back to the properties of the supersymmetric gauge theory entering the construction. This is a six-dimensional YM theory with matter. In \rcite{Chiodaroli:2015rdg}, the matter content of the theory consists of a half-hypermultiplet in a pseudo-real representation of the gauge group. We recall that this choice is made possible by the fact that in six dimensions we can find a constant matrix $B$ obeying the conditions $BB^*=-1$, $[\Gamma^*,B]=0$, where $\Gamma^*$ is the chirality matrix in six dimensions. The fermions in the half-hypermultiplet obey conditions of the form 
\begin{equation}
 \Psi^*= B V \Psi \ , \qquad \Gamma^* \Psi = \Psi \ ,
\end{equation}
where $V$ is a constant antisymmetric matrix with gauge-group-representation indices whose existence is linked to the representation being pseudo-real. With this choice, one can write a kinetic term for the half-hypermultiplet fermions without having to complete the matter content to that of a full hypermultiplet. 
It is however impossible to write a simple mass term without adding extra fields. This would have the form
\begin{equation}
m \Psi^T C V \Psi \ ,
\end{equation}
where $C=\Gamma^0 B$ is the six dimensional charge conjugation matrix. In six dimensions, it is possible to take $C$ to be symmetric, as required for the mass term not to vanish. However, with this choice, $[C,\Gamma^*]\neq 0$ and we need to include spin-$1/2$ fields of both chiralities to have a non-vanishing mass term. 
Hence, it becomes easier to just have matter fields in complex representations. Since this family of constructions is a subset of  the one obtained using pseudo-real representations, we list the supergravities that can be constructed in this way in Table \ref{tabcomplex}.

\begin{table}[t]
\begin{center}
\begin{tabular}{@{\hskip 0.2in}c@{\hskip 0.2in}|@{\hskip 0.2in}c@{\hskip 0.4in}c@{\hskip 0.4in}c}
\bf $D$ & \bf Supergravity & $q$ &$P$ \\[3pt]
\hline
$5$ & Generic non-Jordan family & $-1$ & $2n_F$ \\[3pt]
$6$ & Generic Jordan family & $0$ & $2n_F$ \\[3pt]
$7$ & Homogeneous supergravity & $1$ & $2n_F$ \\[3pt]
$8$ & Homogeneous supergravity & $2$& $n_F$ \\
 &     (Complex Magical sugra when $n_F=1$) && \\[3pt]
$9$ & Homogeneous supergravity & $3$ & $n_F$ \\[3pt] 
$10$ & Homogeneous supergravity & $4$& $n_F$ \\ 
 &     (Quaternionic Magical sugra when $n_F=1$) && \\
 $\vdots$ & $\vdots$  & $\vdots$ & $\vdots$\\
$14$  & Homogeneous supergravity & $8$&$  2n_F$ \\[3pt]
\end{tabular}
\end{center}
\caption{Homogeneous supergravities constructed as double copies involving complex gauge-theory representations. $D$ and $n_F$ are the dimension in which the non-supersymmetric gauge theory can be written and the number of $D$-dimensional fermions. Identification with the parameters $q,P$ from ref. \cite{deWit:1991nm} is provided. Note that, in case of a complex gauge-group representation, chirality conditions can be imposed on the fermions, but not reality conditions.\label{tabcomplex}}
\end{table}

More concretely, we will take all external gauge-theory states with incoming momenta, and label them with their representation. The following fermion polarizations will be used for the supersymmetric gauge theory
\begin{equation}
u=|{\bm p}^{\dot a} \rangle_R  \ ,
	 \qquad 
	 \qquad 
\bar v = \langle {\bm p}^{\dot a} |_{\bar R}	 \ .
\end{equation}
Even if we are using angle spinors for both bra and ket, the sign of the mass is taken to be aligned with the representation. In this way, amplitudes are nonzero provided that the condition $\sum_i m_i=0$ is obeyed. 
It should be stressed that, in this case, $u$ and $\bar v$ are not the conventional particle/antiparticle polarizations, and obey a modified reality condition, as explained in the appendix. 

The states above are combined with the ones from the non-supersymmetric theory; these states will in general carry an extra global index. In some of the cases considered here, the mass matrix in the non-supersymmetric theory has both negative and positive mass eigenvalues. If we diagonalize the mass matrix and denote by $\Sigma$ and $\Sigma'$ the indices corresponding to  eigenvalues of opposite sign, we obtain the states
\begin{equation}
[ {\bm p}^{\dot a}_\Sigma |_{R}  \ ,
 \qquad
\langle {\bm p}^{ a}_{\Sigma'}|_{R} \ ,
\qquad
|{\bm p}_{\Sigma}^{\dot a} ]_{\bar R} \ ,
\qquad
|{\bm p}_{\Sigma'}^{a} \rangle_{\bar R}
\ .
\end{equation}

The global indices  can be combined to obtain a spinor index with respect to the $SO(D-5)$ group coming from the compact dimensions as $(\Sigma, \Sigma')$ . The double copy between two fermions leads to the states 
\bea
W^{a \dot a}_\Sigma &=& |{\bm p}^{a} \rangle_{R} \otimes | {\bm p}^{\dot a}_{\Sigma} ]_{\bar R}    \ ,
\qquad
B^{a  b}_{\Sigma'} = |{\bm p}^{ a} \rangle_{R} \otimes |{\bm p}^{b}_{\Sigma'}\rangle_{\bar R}       \ ,
\no \\
\overline W^{a \dot a}_\Sigma &=&   [{\bm p}^{\dot a}_{\Sigma}|_{R}	\otimes \langle{\bm p}^{ a} |_{\bar R}  \ ,
\qquad
\bar B^{ a  b}_{\Sigma'} = \langle{\bm p}^{ a}_{\Sigma'} |_{R} \otimes \langle{\bm p}^{ b} |_{\bar R}    \ .
\label{vectors_vs_tensors}
\eea
Note that pairing conjugate representations under the double copy prevents us from generating twice as many states. 
For the complex magical theory we have $(\Sigma, \Sigma')= 1,2$, whereas the quaternionic theory has $(\Sigma,\Sigma')=1,2,3,4$.  In some of the above theories, different choices for the individual ranges of $\Sigma$ and $\Sigma'$ are possible, leading to theories with different numbers of vectors and tensors. 
The fermion bilinears above can be expanded as
\begin{eqnarray}
|{\bm p}^{ a} \rangle_{R} \otimes | {\bm p}^{\dot a}_\Sigma ]_{\bar R}  &=& 
-{m_W \over \sqrt{2}} (\cancel{\varepsilon}_{W\Sigma}^{a \dot a} \Omega^{-1}) +{1 \over \sqrt{2}} (\cancel{p} \cancel{\varepsilon}_{W \Sigma}^{a \dot a}\Omega^{-1}) \ ,
\no \\
{[}{\bm p}^{\dot a}_\Sigma |_{R} \otimes \langle {\bm p}^a |_{\bar R}  &=& 
-{m_{\overline{W}} \over \sqrt{2}} (\Omega \cancel{\varepsilon}_{\overline{W} \Sigma}^{a \dot a}) - {1 \over \sqrt{2}} (\Omega \cancel{p} \cancel{\varepsilon}_{\overline{W} \Sigma}^{a \dot a}) \ ,
\end{eqnarray}
which can also be rewritten as
\begin{eqnarray}
|{\bm p}^{ a} \rangle_{R} \otimes [ {\bm p}^{\dot a}_\Sigma |_{\bar R}  &=& 
{m_W \over \sqrt{2}} (\cancel{\varepsilon}_W)^{a \dot a}_\Sigma -{1 \over \sqrt{2}} \cancel{p} (\cancel{\varepsilon}_W)^{a \dot a}_\Sigma \ ,
\no \\
|{\bm p}^{\dot a}_\Sigma ]_{R} \otimes \langle {\bm p}^a |_{\bar R}  &=& -
{m_{\overline{W} } \over \sqrt{2}} (\cancel{\varepsilon}_{\overline{W}})^{a \dot a}_\Sigma - {1 \over \sqrt{2}} \cancel{p} (\cancel{\varepsilon}_{\overline{W}})^{a \dot a}_\Sigma \ .
\end{eqnarray}
In the above equation, $W^{a \dot a}_\Sigma$ are states corresponding to a massive W-boson transforming in a complex representation of the  unbroken supergravity gauge group, while $\overline W^{a \dot a}_\Sigma$ give the states of its conjugate. 

The tensor bilinears obey a similar relation
\begin{equation}
|{\bm p}^{a} \rangle_{R} \otimes \langle {\bm p}^b_\Sigma |_{\bar R}  =  {1 \over 2} \epsilon^{a  b} ({\cancel p + m }) -
{m \over \sqrt{2}} (\cancel{\varepsilon}_B)^{ a b}_{\Sigma'} \ .
\end{equation}
The above construction with complex representations provides the correct number of degrees of freedom for the complex and quaternionic magical theories. However, a naive extension of the construction for the real and octonionic theories presented in \rcite{Chiodaroli:2015wal} to complex representations yields too many degrees of freedom (in the case of the octonionic theory a total of 32 vectors or tensors are constructed from fermion bilinears, as opposed to 16). It is possible to check that this double copy gives the homogeneous $L(q,P)$ theories with $q=1$ and $P=2$ and  $q=8, P=2$ instead of the real and octonionic magical theory ($q=1$, $P=1$ and $q=8$, $P=1$). Obtaining these theories from double-copy constructions with complex fields requires additional projections, as it will be discussed in the conclusion.

The construction outlined above must be further generalized if the spectrum of the theory contains more than one nonzero mass (up to a sign). A possible approach is to assign different gauge-group representations to matter hypermultiplets with different masses. Examples of this generalized construction will be discussed in later sections of this paper.

\subsection{Three-point amplitudes from the double copy: the dictionary}

We now study how amplitudes obtained from the double copy yield supergravity interactions of the types discussed in Section~\ref{ampgrav}.
A particularly relevant subclass of amplitudes we shall consider is amplitudes between two massive and one massless vector, since they 
probe the spontaneously-broken part of the supergravity gauge group.\footnote{We include in this class massive vector fields associated with non-compact gaugings since in their vacua the non-compact gauge group is broken to its maximal compact subgroup.} 
A massive vector can arise both as a bosonic double copy of one massive vector and one massive scalar from the gauge theories and as a fermionic double copy of two spin-$1/2$ fields (with opposite sign of the masses).

\def\sugA{V}

In the case of a bosonic double copy we have three distinct possible three-vector amplitudes. The first has the form\footnote{We denote as $\sugA_\mu$ massless supergravity vectors obtained as double copies, to differentiate them from the fields from the supergravity Lagrangian.}
\bea
{\cal M}_3(1 \overline W,2W, 3 \sugA)  &=& i A_3(1 \overline W,2W, 3A)    A_3(1 \bar \varphi, 2\varphi, 3\phi)\,  \\
&=& - { \lambda \over 2 } \Big( \langle \overline W W A \rangle \Big|_{|D W|^2} + \langle  \overline W W A \rangle \Big|_{\overline{W} \cdot F \cdot {W} } \Big)
= -
{ \lambda \over 2 } \langle \overline W W A \rangle \Big|_{\rm YM} , \nn
\eea
where $\lambda$ is the constant appearing in the three-scalar coupling. The second class of amplitudes differs because the massive and massless scalars come from different gauge theories,
\bea
{\cal M}_3(1 \overline W,2W , 3\sugA)  &=& i A_3(1 \overline W,2W, 3\phi)  A_3(1 \bar \varphi, 2\varphi, 3A) \nn \\
&=& i { m }  ( \bep_1 \cdot \bep_2) ( \varepsilon_3 \cdot p_1 - \varepsilon_3 \cdot p_2)  \nn \\
&=&    m  \langle \overline W W A \rangle_{|D W|^2} -    m  \langle  \overline W W A \rangle_{{\overline{W} \cdot F \cdot {W} } } \ . \nn
\eea
The last possibility for a bosonic double copy has the massless vector constructed as an antisymmetric combination of the two gluons from the gauge theories
\bea \label{WWA2}
{\cal M}_3(1 \overline W,2 W, 3 \sugA)& =& -i {\epsilon^{ab}\over 4} {\partial \over \partial \auxu_3^{a}}   A_3(1 \overline W,2W, 3A)  {\partial \over \partial \auxu_3^{b}}   A_3(1\bar \varphi,2\varphi, 3A)  \nn \\
&=&  - {1 \over 2 \sqrt{2}} \langle  \overline W W A \rangle_{  D\overline{W} \wedge A \wedge DW} \,.
\eea
In this case, the amplitude is the same as in the massless case. 

If the massive vectors are realized as fermionic double copies, we have two possible three-point amplitudes. The first involves the massless vector is constructed as the antisymmetric combination of the gauge-theory bosons:
\bea
{\cal M}_3(1 \overline W,2W, 3 \sugA)  &=&- i {\epsilon^{ab} \over 4} {\partial \over \partial \auxu_3^{a}} A_3(1 \bar \chi,2  \chi, 3A) {\partial \over \partial \auxu_3^{b}} A_3(1 \bar \chi,2  \chi, 3A) \nn \\
&=&{ \sqrt{2}}i m \Big[ (\bep_1 \cdot \varepsilon_3 \, p_3 \cdot \bep_2 -\bep_2 \cdot \varepsilon_3 \, p_3 \cdot  \bep_1  )  \Big] \nn \\
&=&  -{i \over 4 \sqrt{2} m}  \Big(\abra{1} | p_2 |\sket{1} \, \abra{2} | \varepsilon_3 | \sket{2}  - (1 \leftrightarrow 2)  \Big)  \\
&=&{ \sqrt{2}} m   \langle W \overline W A \rangle_{\overline{W} \cdot F \cdot {W} } \ . \nn
\label{fermifermi_v1}
\eea
Fermionic masses are taken as $m_1=-m$, $m_2=m$.
We see that this particular vector couples to other massive vectors obtained from fermionic bilinears through its field strength, but does not appear in the covariant derivatives. As we have previously mentioned, this type of interactions are characteristic to gauge groups with a central charges. 
Since such a massless vector is ubiquitous to the double-copy framework, many of the gaugings we construct in this work will have the form $G\ltimes {\cal H}$, where $\cal H$ is the Heisenberg group.

A second class of double-copy amplitudes involving supergravity vectors constructed as fermion bilinears is
\bea
{\cal M}_3(1\overline W,2W, 3 \sugA)  &=&i  A(1 \bar \chi,2  \chi, 3A) A(1 \bar \chi,2   \chi, 3\phi)   \\
&=&  {i \over 2} \sbra{2}\sket{1} \,  \abra{1} | \varepsilon_3 | \aket{2} 
= { m}  \langle \overline W  W A \rangle_{|D W|^2}-  {1 \over 4}\langle  \overline W W A \rangle_{D\overline{W} \wedge A \wedge DW } , \nn \qquad
\label{fermifermi_v2} 
\eea
where we have used again $m_1=-m, m_2=m$. In contrast to the previous case, we see that covariant-derivative couplings arise on top of the interactions already present in the ungauged theory.

As discussed, the double-copy realization of massive tensors is closely related to that of massive vectors, cf. eq.~\eqref{vectors_vs_tensors}. We may therefore obtain three-point amplitudes with one massless vector and  two massive tensors by simply rearranging the order of fields in eqs.~\eqref{fermifermi_v2} and ~\eqref{fermifermi_v1}.
If the massless vector is realized as a scalar-vector pair, the amplitude is 
\begin{eqnarray}
{\cal M}_3\big( 1 \bar B,2 B,3 \sugA \big) &=&   i A_3\big( 1 \bar \chi ,2 \chi ,3 A \big) A_3\big( 1 \bar \chi ,2 \chi ,3 \phi^i \big) \no\\
&=& {i \over 2}   \abra{1}\aket{2} \,  \abra{2} | \varepsilon_3 | \aket{1} \ 
 = - i m^2 \epsilon\big(\varepsilon^B_1, {\varepsilon}_3,  \varepsilon^B_2 \big) \ .
\end{eqnarray}
Note that, from the supergravity perspective, this amplitude comes entirely from the $B\wedge B \wedge A$ term. 
Finally, if the special vector is a double copy of two gluons the three-point amplitude vanishes,
\bea
{\cal M}_3(1 \bar B,2B, 3 \sugA)  &=&- i {\epsilon^{ab} \over 4} {\partial \over \partial \auxu_3^{a}} A_3(1 \bar  \chi,2  \chi, 3A) {\partial \over \partial \auxu_3^{b}} A_3(1 \bar \chi,2  \chi, 3A) = 0
 \ .
\eea
The amplitudes between vectors and tensors presented here give the fundamental ingredients for analyzing the output of several possible double copies. In general, fields in the two gauge theories will have extra indices that require a case-by-case study, and result in different options for the choice of the supergravity gauge group. C/K duality typically constrains the couplings that appear in the gauge theories as well as the matter content and spectra, while matching of the 
supergravity spectra also requires a careful analysis.  The basic kinematic dependence of the amplitudes is however the same in all cases, and the dictionary established in this section will be consistently used in the remainder of the paper. 

\subsection{Summary of the constructions presented in this paper}

We are now ready to summarize the main results in this paper. Concretely, different constructions will depend on the number of scalar fields in the non-supersymmetric gauge-theory factor, the choice for the $F^{ijk}$ tensors, as well as the choice of representation for the fermions, which will also affect how fields are paired by the double copy. Details about individual constructions and, in particular, matching with the supergravity Lagrangians, will be given in the following sections.  
\begin{enumerate}
\item 
Non-compact $SO(2,1)$ gauging for the generic Jordan family. This construction has the schematic form
\begin{equation}
\Big(\cN=2\text{ SYM on Coulomb branch} \Big) 
\otimes
\Big( \text{YM  + $n_s$ massive scalars}      \Big) \ .
\end{equation}
This construction always yields two massive gauge fields. At least one massless scalar is present, but additional massless scalars can also be added. Specific choices for the trilinear couplings between two massive and one massless scalars can yield either a non-compact $SO(2,1)$ gauging or a compact $SO(3)$ gauging. 

\item
Generic Jordan family, $U(1)\ltimes \mathcal{H}_{2P+1}$ gauging. This has the construction:
\begin{equation}
\Big( \cN=2\text{ SYM  + hyper}{}_R \Big) 
\otimes
\Big( \text{YM  + 1 scalar + $P$ massive fermions}{}_{\bar R}   \Big) \ .
\end{equation}
The appearance of a central charge seems to be a generic feature of the double copy and is related to the presence of a massless vector field in the spectrum that originates from the double copy of two gauge-theory gluons.

\item 
Generic Jordan family, $U(1)$ gauging with massive tensors. This is a variant of the previous construction in which different signs for the fermionic masses are used.

\item 
Complex theory, $U(2)\ltimes \mathcal{H}_5$ gauging.
Schematically, the 5D double-copy construction has the form 
\begin{equation}
\Big( \cN=2\text{ SYM  + hyper}{}_R \Big) 
\otimes
\Big( \text{YM  + 3 scalars + 2 massive fermions}{}_{\bar R}   \Big) \ ,
\end{equation}
where the $F^{ijk}$ tensor determining the fermion mass matrix, cf. eq.~\eqref{fermionmass}, is chosen to give the unbroken $U(2)$ gauge symmetry. This construction yields at least four massive vectors, but taking the first gauge-theory factor on the Coulomb branch and adding masses for some of the scalars results in a double-copy theory in which $U(2)$ is spontaneously broken and hence two additional vectors acquire masses.  These gaugings can also be seen as $U(2,1)$ gaugings in which a pp-wave limit has been taken.

\item Complex magical theory, $U(2)$ gauging with massive tensors. This is obtained as a seemingly minor variant of the previous construction in which matter representations are chosen so that tensors arise instead of vectors in the double copy.
This construction yields four massive tensors, but taking the first gauge-theory factor on the Coulomb branch and adding masses for some of the scalars results in a double-copy theory in which $U(2)$ is spontaneously-broken.  

\item Another variant of the above construction has four massive vectors, two massive tensors and three massless vectors. It corresponds to a 
$U(1,1) \ltimes \mathcal{H}_3$ gauging of the complex theory.

\item 
Quaternionic theory,  $U(2) \ltimes {\cal H}_5$ gauging. This construction has the schematic form
\begin{equation}
\Big( \cN=2\text{ SYM  + hyper}{}_R \Big) 
\otimes
\Big( \text{YM  + 5 scalars + 4 massive fermions}{}_{\bar R}   \Big) \ .
\end{equation}
All the vector fields in the quaternionic theory transform in the adjoint representation of the $SO^*(6)=SU(3,1)$ subgroup of its global symmetry group $SU^*(6)$. 
{
The group $SU(3,1)$ leads to the group $SU(3)\ltimes \mathcal{H}_7 $ by taking a particular pp-wave limit. One can gauge the $[U(2)\ltimes \mathcal{H}_5 ]  $ subgroup of $SU(3)\ltimes \mathcal{H}_7 $. } 
The spectrum then  contains four  massive vector fields, four  massive tensors and a  vector field corresponding to the central charge.  In addition to these, there are two  extra massless vectors that partake in couplings between massive vectors and massive tensors. This indicates that the construction gives a gauging with an abelian invariant subalgebra along the lines of \rcite{Bergshoeff:2004kh}.  
One can also spontaneously break $U(2)$ to a $U(1)^2$ subgroup and obtain two additional massive vectors in the spectrum.  

\item Another variant of the above construction with 6 massive vectors is given by gauging  $U(1,1) \ltimes {\cal H}_5$ symmetry. This realization can be truncated to the $U(1,1) \ltimes {\cal H}_3$ gauging for the complex theory discussed above. 

\end{enumerate}

\noindent
We also comment on the construction
\begin{equation}
\Big( \cN=2\text{ SYM  + hyper}{}_R \Big) 
\otimes
\Big( \text{YM  + 9 scalars + 8 massive fermions}{}_{\bar R}   \Big) \ , \no
\end{equation}
which gives gaugings for the homogeneous theory with $q=8,P=2$, and mention in the conclusion  the possibility of obtaining the octonionic theory through a 
truncation after the double copy.

\section{Gaugings for the generic Jordan family \label{sec4}}

\subsection{Gauge-theory factors and double-copy amplitudes}

The supersymmetric (left) theory is a spontaneously-broken theory similar to the ones studied in \rcite{Chiodaroli:2015rdg}. We only write the bosonic part of the Lagrangian,
\begin{eqnarray}
{\cal L}_{\rm L} &=&
-{1 \over 4} \Big( F_{\mu \nu}^{\hat a} +2 i g  \overline{W}_{[\mu} t^{\hat a} W_{\nu]} \Big)^2 +
{1 \over 2} \Big(D_\mu \phi^{\hat a}  +
i \overline{W}_\mu t^{\hat a}  \varphi-i\overline{\varphi} t^{\hat a}  W_\mu   \Big)^2
 \no \\
&&  \qquad -2 \big|D_{[\mu}W_{\nu]}\big|^2  + \big| D_\mu \varphi +i m W_\mu + i \phi^{\hat a} t^{\hat a} W_\mu   \big|^2 + \text{fermions} \ .
\end{eqnarray}
This theory contains massless adjoint gluons $A^{\hat a}_{\mu}$ and scalars $\phi^{\hat a}$, the former appearing through their respective field strengths $F^{\hat a}_{\mu \nu}$. $\hat a, \hat b  \ldots$ are adjoint indices of the unbroken gauge group.

The massive non-adjoint fields  of the theory consist of $W_\mu$ and $\overline{W}_\mu$ bosons, transforming in conjugate representations $R$ and $\bar R$  of the unbroken gauge group, and their respective multiplets.  Hermitian representation matrices are denoted as $t^{\hat a}$. The scalar fields $\varphi$ and $\bar \varphi$ are eaten by the corresponding massive vector fields, and decouple manifestly from the theory when an appropriate choice of the gauge-fixing term is made. 
A theory of this sort can be obtained from an unbroken, purely adjoint gauge theory 
with $SU(N_1+N_2)$ gauge group giving the five-dimensional scalar field a vev
breaking the gauge group to $SU(N_1)\times SU(N_2) \times U(1)$. 
In this case, the structure constants of the original gauge group give the structure constants of the unbroken gauge group $f^{\hat a \hat b \hat c}$ and the representation matrices $t^{\hat a}$. This implies that, aside from the standard commutation relations, the representation matrices obey color relations stemming from the Jacobi relation of the original theory before the vev is introduced. These extra relations have the form 
\begin{equation}
(t^{\hat a})^{ \ \hat  \beta}_{ \hat \alpha} (t^{\hat a})^{ \ \hat  \delta}_{\hat  \gamma} =  (t^{\hat a})^{ \ \hat  \delta}_{ \hat  \alpha} (t^{\hat a})^{ \ \hat  \beta}_{ \hat  \gamma} \label{extra} \ ,
\end{equation}
where we have now explicitly displayed the representation indices. This is a particular case of the seven-term relation in \rcite{Chiodaroli:2015rdg}. 
There are only two  amplitudes with one massless  and two massive fields in five dimensions:\footnote{Partial amplitudes are normalized as ${\cal A}(1,2,3)= g {\rm Tr}(T^{\hat a_1} T^{ \hat a_2} T^{\hat a_3}) A(1,2,3)$, with Tr$(T^{\hat a_1} T^{\hat a_2})= \delta^{\hat a_1 \hat a_2}$, while generators appearing in the Lagrangian have the standard textbook normalization Tr$(t^{\hat a_1} t^{\hat a_2})= {1 \over 2} \delta^{\hat a_1 \hat a_2}$. }
\begin{eqnarray}
A_3 \big( 1 \overline{W} , 2 W, 3 \phi \big) &=&  i \sqrt{2} m (\varepsilon_1 \cdot \varepsilon_2)  \ , \nn \\
A_3 \big( 1 \overline{W}, 2 W, 3 A \big) &=&  i \sqrt{2}   (k_2 \cdot \varepsilon_1) (\varepsilon_2 \cdot \varepsilon_3) + \text{cyclic} = {1 \over \sqrt{2}} \langle \overline{W} W A  \rangle_{\rm YM}   \ .
\end{eqnarray}

We now turn to the non-supersymmetric (right) theory. We choose a YM theory
 with one massless adjoint scalar $\phi$ and a complex massive scalar in the conjugate representations
to the one of the $W$ multiplet in the left gauge theory. 
We also introduce a trilinear coupling between massless and massive scalars.
The full Lagrangian is 
\begin{eqnarray}
\cL_{\rm R} &=& - {1 \over 4} (F^{\hat a}_{\mu \nu})^2  + {1 \over 2} (D_{\mu} \phi^{\hat a})^2  + | D_{\mu}\varphi|^2  - m^2  |\varphi|^2 \no + g \lambda \phi^{\hat a} \bar \varphi t^{\hat a} \varphi
\\
&& \qquad
-{g^2 \over 2} (\bar \varphi t^{\hat a}  \varphi ) (\bar \varphi t^{\hat a}  \varphi ) - g^2 \phi^{\hat a} \phi^{\hat b} (\bar \varphi t^{\hat a}t^{\hat b} \varphi ) \ .
\end{eqnarray}
This theory obeys C/K duality. In particular, inspecting the four-scalar amplitude 
\begin{displaymath}
\cA_4 \big( 1 \bar{\varphi}^{\hat \alpha}, 2 \bar{ \varphi}^{\hat \beta}, 3 \varphi_{\hat \gamma}, 4  \varphi_{\hat \delta} \big)
\end{displaymath}
leads to the numerator and color factors
\begin{equation}
n_u = n_t = 2s + \lambda^2 \ , \qquad c_u = (t^{\hat a})_{ \hat \gamma}{}^{\hat \alpha} (t^{\hat a})_{\hat \delta}{}^{\hat \beta} \ , \qquad
\qquad c_t = (t^{\hat a})_{ \hat \delta}{}^{\hat \alpha} (t^{\hat a})_{\hat \gamma}{}^{\hat \beta} \ ,
\end{equation}
and the extra identity (\ref{extra}) is satisfied by the numerator factors. 
The only relevant three-point amplitudes with two massive and one massless fields for the non-supersymmetric gauge theory are the following:
\begin{eqnarray}
A_3 \big( 1 \overline \varphi, 2 \varphi, 3 A \big) &=&  - i\sqrt{2}  (p_1 \cdot \varepsilon_3)  \ , \nn \\
A_3 \big( 1 \overline \varphi, 2 \varphi, 3 \phi \big) &=&  {i \over \sqrt{2}} \lambda
\end{eqnarray} 
Looking at the output of the double copy, and using the results from the previous subsections, we have three distinct supergravity amplitudes between one massless and two massive   vectors\footnote{Here the normalization of the double-copy follows the KLT formula with the gravitational coupling set to $\kappa=2$, {\it i.e.} ${\cal M}_3 =  {i} A(1,2,3) \tilde A(1,2,3)$.} 
\begin{eqnarray}
\cM_3 \big( 1 \overline W, 2 W , 3 \sugA^0 \big) &=&  - {1 \over 2 \sqrt{2}}    \langle  \overline W W A \rangle_{D\overline{W}  \wedge DW \wedge A }  \ , \no \\
\cM_3 \big( 1 \overline W, 2 W , 3 \sugA^1\big) &=& 2 i m (p_1 \cdot \varepsilon_3)  (\varepsilon_1 \cdot \varepsilon_2) = m \Big( \langle \overline W W A \rangle_{|D W|^2} -      \langle  \overline W W A \rangle_{{\overline{W} \cdot F \cdot {W} } }\Big) \ , \no \\
\cM_3 \big( 1 \overline W, 2 W , 3 \sugA^4 \big) &=& -i \lambda  \big( (k_2 \cdot \varepsilon_1) (\varepsilon_2 \cdot \varepsilon_3) + \text{cyclic} \big)  = - {\lambda \over 2}  \langle \overline W W A \rangle_{\rm YM} \ .
\end{eqnarray}
Here $\sugA^0$ denotes a vector obtained from the double copy of two gauge-theory gluons, and $\sugA^1$ and $\sugA^4$ are two distinct massless vectors arising from the double copies of adjoint vectors and scalars from the two sides of the theory.

\subsection{Supergravity amplitudes\label{Jordan_conventional}}

To identify the supergravity Lagrangian for the theory obtained from the double copy, we consider the example of a $SO(2,1)$ gauging of theories belonging to the generic Jordan family in five dimensions. To fix notation, we take the cubic polynomial 
\begin{equation}
{\cubpol}\big(\xi^I\big) = \sqrt{2} \xi^0 \big( {(\xi^1)}^2 - {(\xi^2)}^2 - \ldots -{(\xi^{n_V})}^2   \big) \ , 
\end{equation}
where $n_V$ is the number of vector multiplets in five dimensions. We take  $n_V\geq 3$ and 
proceed to identify the relevant non-compact transformations written in ambient space according to eq. (\ref{transformations}).
The $SO(2,1)$ generators $T$ act nontrivially only in the $(\xi^1, \xi^2, \xi^3)$ subspace. They are
\begin{equation}
T_1 = \left( \begin{array}{ccc} 0 & 0 & 0 \\ 0 & 0 & -1 \\ 0 & 1 & 0 \end{array} \right) \ ,
\quad
T_2 = \left( \begin{array}{ccc} 0 & 0 & 1 \\ 0 & 0 & 0 \\ 1 & 0 & 0 \end{array} \right) \ ,
\quad
T_3 = \left( \begin{array}{ccc} 0 & 1 & 0 \\ 1 & 0 & 0 \\ 0 & 0 & 0 \end{array} \right) \ ,
\end{equation}
Note that the compact generators are anti-hermitian, while the non-compact 
generators are hermitian.
The structure constants for this choice of generators are 
\begin{equation}
f^3{}_{12} = -1  \ , \qquad f^2{}_{31} = -1 \ , \ \qquad f^1{}_{23} = 1 \ .
\end{equation}
Amplitudes are obtained expanding the Lagrangian around the base-point\footnote{One can consider the more general choice $c^I = \big( e^{-2\theta}/\sqrt{2}, \ e^\theta \cosh \phi, \ 0 , \ 0,  \ e^\theta\sinh{\phi},   \ldots \big) $, but the extra parameter $\theta$ can be absorbed by a redefinition of $\gsug$.} 
\begin{equation}
\xi^I=c^I = \big( 1/\sqrt{2}, \  \cosh \phi, \ 0 , \ 0,  \ \sinh{\phi},  \ 0 , \ldots \big) \ .
\end{equation}
With this choice, the $(n_V+1)\times (n_V+1)$ matrix that appears in the kinetic term for the vector fields is
\begin{equation}
\å_{IJ} \big|_0 = \left( \begin{array}{ccccccc}
1 & 0 & 0 & 0 & 0  &\cdots& 0 \\ 
0 & \cosh 2\phi & 0 & 0 & -\sinh 2\phi  &\cdots& 0 \\ 
0 & 0 & 1 & 0 & 0  &\cdots& 0 \\
0 & 0 & 0 & 1 & 0  &\cdots& 0 \\
0 & -\sinh 2 \phi & 0 & 0 & \cosh 2\phi  &\cdots& 0 \\
\vdots & \vdots  & \vdots &  & \vdots  & &  \vdots \\ 
0 & 0 & 0 & 0 & 0  &\cdots& 1 \\    
\end{array} \right) \ .
\end{equation} 
The $n_V \times n_V$ matrix in the kinetic term for the scalars is 
\begin{equation}
g_{xy} \big|_0 = \left( \begin{array}{cccccc} 
 1 + 2\cosh 2\phi & 0 & 0 & -2\sinh 2\phi  &\cdots& 0 \\ 
  0 & 1 & 0  &0&\cdots& 0 \\
 0 & 0 & 1 & 0  &\cdots& 0 \\
 -2\sinh 2 \phi & 0 & 0 & 2\cosh 2\phi - 1  &\cdots& 0 \\
\vdots  & \vdots &  &  \vdots  & &  \vdots\\ 
 0 & 0 & 0 & 0  &\cdots& 1 \\    
\end{array} \right) \ .
\end{equation} 
Explicit calculation with our choice of base-point leads to
\begin{equation}
M^2_{IJ} = \gsug^2 \, \text{diag} \Big( 0, 0, \cosh^2 \phi,  \cosh^2 \phi, 0 ,\ldots \Big)
\label{massSQmatrix}
\end{equation}
for the vector fields. It is then convenient 
to redefine the fields so that the vector-field kinetic term becomes canonically normalized,
\begin{eqnarray}
 A^1_\mu &\rightarrow& \cosh \phi \, {A}^1_\mu + \sinh \phi \, {A}^4_\mu \ , \nn  \\
 A^4_\mu &\rightarrow& \sinh \phi \, {A}^1_\mu + \cosh \phi \, {A}^4_\mu  \ ,
\end{eqnarray}
where the other fields are left invariant. 
The final result for the relevant three-vector amplitudes after this field redefinition is
\begin{eqnarray}
{\cal M}_3(1A^2,2A^3,3A^1) &=& -2 \gsug \cosh \phi  \,   (p_1\cdot \varepsilon_3) (\varepsilon_1\cdot \varepsilon_2) = -2 m   (p_1\cdot \varepsilon_3) (\varepsilon_1\cdot \varepsilon_2) , \nn \qquad \\
{\cal M}_3(1A^2,2A^3,3A^4) &=& i \gsug \sinh \phi \, \langle  \overline W W A \rangle_{\rm YM}.  \label{threeampso21}
\end{eqnarray}
Note that now $A^2$ and $A^3$ are massive, leading to the identification with the double-copy fields 
\begin{equation}
A^0_\mu=V^0_\mu\ , \qquad A^1_\mu=V^1_\mu\ , \qquad  A^4_\mu=V^4_\mu\ ,  \qquad  W_\mu={1 \over \sqrt{2}} (A^2_\mu+iA^3_\mu) \ . 
\end{equation}
We see that two parameters are accessible at the level of the amplitudes from the supergravity Lagrangian: the supergravity coupling constant $\gsug$ and the angle $\phi$. 

These Lagrangian-based amplitudes can be matched by the double-copy calculation with the identification 
\begin{equation}
\lambda = -2 \gsug \sinh \phi \ .
\end{equation}
Since $m=\gsug \cosh \phi$, the non-compact gauging implies the relation $|\lambda| < 2  m$. 
The structure constants corresponding to the amplitudes above are
\begin{eqnarray}
f^1_{\ 23}= \cosh \phi \ , \qquad f^2_{\ 31}=  -\cosh \phi \ , \qquad f^3_{\ 12}= - \cosh \phi \ , \no \\
f^4_{\ 23}= -\sinh \phi \ , \qquad f^2_{\ 34}= -\sinh \phi \ , \qquad f^3_{\ 42}=  -\sinh \phi \ .
\end{eqnarray}
They correspond to the commutation relations between generators
\begin{eqnarray}
 [ t_2,t_3]&=&  \cosh \phi \ t_1 - \sinh \phi \ t_4 \ , \nn \\
{ [ t_1,t_3]}&=&  \cosh \phi \ t_2 \ , \nn\\ 
{ [ t_1,t_2]}&=& - \cosh \phi \ t_3  \ , \\
{[t_3,t_4]}&=& -\sinh \phi \ t_2 \ , \nn \\
{[t_2,t_4]}&=&  \sinh \phi \ t_3 \ .\nn
\end{eqnarray}
For $\phi$ finite, defining the new generators $\cosh \phi \, t_1 -\sinh  \phi  \,t_4$ and $ \sinh \phi \, t_1 - \cosh \phi \, t_4$, we can see that the group in indeed isomorphic to $SO(2,1)$, where the second generator decouples. However, for $\phi \neq 0$, the above field redefinition does not preserve the  diagonalized form of the  kinetic term for the vector fields at the base-point. Hence, it is not possible to decouple the field $A^4_\mu$ in eq. (\ref{threeampso21}). We also note that the physical graviphoton ({\it i.e.} the combination appearing in the supersymmetry transformation of the graviton) is given by
\begin{equation}
{1 \over \sqrt{2}} A^0_\mu + A^1_\mu \ ,
\end{equation}
while the field $A^0$ couples to the others only through the $C_{IJK}$ term in the action, which is unaffected by the gauging procedure. 

We can repeat the analysis for the compact $SO(3)$ gauging, using the vectors $A^2_\mu, A^3_\mu, A^4_\mu$.
The base-point is left as before while 
the matrices $T$ generating $SO(3)$ now act nontrivially only in the $(\xi^2, \xi^3, \xi^4)$ subspace,
\begin{equation}
T_1 = \left( \begin{array}{ccc} 0 & 0 & 0 \\ 0 & 0 & -1 \\ 0 & 1 & 0 \end{array} \right) \ ,
\quad
T_2 = \left( \begin{array}{ccc} 0 & 0 & -1 \\ 0 & 0 & 0 \\ 1 & 0 & 0 \end{array} \right) \ ,
\quad
T_3 = \left( \begin{array}{ccc} 0 & -1 & 0 \\ 1 & 0 & 0 \\ 0 & 0 & 0 \end{array} \right) \ ,
\end{equation}
where all other entries are zero. After repeating a similar computation with respect to the non-compact gauging, we get a mass matrix of the form
\begin{equation}
M^2_{IJ} = \gsug^2 \text{diag} \Big( 0, 0, \sinh^2 \phi,  \sinh^2 \phi, 0 ,\ldots \Big) \ , 
\end{equation}
where one can see that setting $\phi=0$ gives the unbroken $SU(2)$ phase.
The final result for the relevant three-vector amplitudes after the field redefinition is
\begin{eqnarray}
{\cal M}_3(1A^2,2A^3,3A^1) &=& -2 \gsug \sinh \phi\,   (p_1\cdot \varepsilon_3) (\varepsilon_1\cdot \varepsilon_2) = -2 m   (p_1\cdot \varepsilon_3) (\varepsilon_1\cdot \varepsilon_2) \nn \\
{\cal M}_3(1A^2,2A^3,3A^4) &=& i \gsug \cosh \phi \, \langle  \overline W W A \rangle_{YM} 
\end{eqnarray}
{\it i.e.} hyperbolic sine and cosine are exchanged. The parameter identification from the double copy is now
\begin{equation}
\lambda =- 2 \gsug \cosh \phi \ ,
\end{equation}
and  the compact gauging solution obeys the condition $|\lambda| > 2m$.  

Note that the solution with $\lambda= \pm 2m$ corresponds to taking the limit $\phi \rightarrow \pm \infty$. In this limit, it is convenient to take the linear combinations $A^1_\mu \pm A^4_\mu$; combined with the massive vectors, they give a gauging of the form 
\begin{equation}
SO(2) \ltimes \mathcal{H}_3 \ ,
\end{equation}
where ${\cal H}_3$ is the three dimensional Heisenberg group. This limit can also be thought of as a pp-wave limit.

The appearance of the Heisenberg group may look surprising at first sight since $SO(3)$ is compact. However both $SO(2,1)$ and $SO(3)$ admit contractions to the Euclidean group $E_2=ISO(2)=SO(2)\ltimes \mathbb{T}_2 $. Interestingly, in the limit $\lambda= \pm 2m$ the translations $\mathbb{T}_2$ close into a central charge in the supergravity theory which is gauged by the extra vector field other than the  graviphoton. 

At this point it may be useful to point out that simple non-compact groups $G$  admit what is called the Iwasawa decomposition, which decomposes any group element $G$  uniquely as\footnote{We are using the same symbol for denoting the group element and the group.}
\be G = K  A  N \ , \ee
where $K$ belongs the maximal compact subgroup, $A$ is an element of some suitable abelian subgroup  and $N$ is an element of a nilpotent subgroup of $G$. \footnote{For a review of Iwasawa decomposition of non-compact groups we refer to ref. \cite{MR1920389}.} Both $A$ and $N$ are closed subgroups of $G$. The nilpotent subgroup $N$ is normalized by $A$. For the group $G=SL(m,\mathbb{C})$ of $m \times m $ complex matrices with unit determinant, the group $K$ is $SU(m)$, $A$ is the Abelian group of diagonal matrices with positive entries and $N$ is the nilpotent group generated by $m\times m $ upper triangular matrices with 1 in each diagonal entry. 
At the Lie-algebra level, we have the corresponding decomposition 
\be \mathfrak{g} = \mathfrak{k} \oplus \mathfrak{a} \oplus \mathfrak{n} \ , \ee
such that  $\mathfrak{g} , \mathfrak{k} , \mathfrak{a}$ and $\mathfrak{n}$ denote the Lie algebras of $G, K, A$ and $N$, respectively. $\mathfrak{a}$ is abelian,  $\mathfrak{n}$ is nilpotent, $\mathfrak{a}\oplus \mathfrak{n}$ is a solvable subalgebra, and  $[\mathfrak{a}\oplus \mathfrak{n}, \mathfrak{a}\oplus \mathfrak{n}]=\mathfrak{n}$.
For studying the non-compact gaugings of  five-dimensional supergravity  theories with non-compact  U-duality groups  $G$, the subgroups $H$ of  the form 
\be 
H=M A N \ ,
\ee
will be particularly important.  $M$ is a subgroup of $K$ which commutes with the abelian subgroup $A$ such that both $M$ and $A$ normalize a nilpotent subgroup $N$. 

\subsection{Another realization for the generic Jordan family}

In \rcite{Chiodaroli:2015rdg} it was noted that the generic Jordan family has two distinct double-copy realizations. Aside from the purely-adjoint double copy described in the previous subsection, we also can consider a construction along the lines of the homogeneous theories in which the right gauge-theory factor is a six-dimensional pure YM theory augmented with $P$ irreducible fermions.
In five dimensions, the supersymmetric (left) gauge theory has a massless vector multiplet and one massive complex hypermultiplet transforming in the fundamental representation; its Lagrangian  is given by 
\begin{eqnarray}
\label{susyLhyper}
{\cal L}_{L}&=& -{1\over 4} (F_{\mu \nu}^{\hat a})^2  + i \bar \psi^{\hat a} \cancel{D} \psi^{\hat a} + {1 \over 2} (D_\mu \phi^{\hat a})^2 +  |D_\mu \varphi|^2 
+ {i } \bar \chi \cancel{D} \chi -  m \bar \chi \chi  -  m^2 | \varphi|^2  \\[1pt]
&& \hskip -1.3cm +\sqrt{2} g \bar \psi^{\hat a} \varphi t^{\hat a} \chi +\sqrt{2} g \bar \chi t^{\hat a} \varphi \psi^{\hat a} + i g f^{\hat a \hat b \hat c} \phi^{\hat a} \bar \psi^{\hat b} \psi^{\hat c}  + g^2  \phi^{\hat a} \phi^{\hat b} \bar \varphi t^{\hat a} t^{\hat b} \varphi
+ g \bar \chi t^{\hat a} \chi \phi^{\hat a} + g^2 (\bar \varphi t^{\hat a} \varphi)^2  \ . 
\no 
\end{eqnarray} 
As usual, the covariant derivatives encode the gauge-group representations of the fields; for our purpose they are given by
\begin{align}
(D_\mu \psi)^{\hat a} =& \partial_\mu \psi^{\hat a} + g f^{\hat a \hat b \hat c} A_\mu^{\hat b} \psi^{\hat a} \ , &\qquad
(D_\mu \phi)^{\hat a} =& \partial_\mu \phi^{\hat a} + g f^{\hat a \hat b \hat c} A_\mu^{\hat b} \phi^{\hat a} \ , \\
(D_\mu \chi) =& \partial_\mu \chi -i g t^{\hat a } A_\mu^{\hat a} \chi \ ,
& \qquad
(D_\mu \varphi) =& \partial_\mu \varphi -i g t^{\hat a } A_\mu^{\hat a} \varphi \ .
\nonumber
\end{align}
The  non-supersymmetric (right) theory  has  Lagrangian
\begin{eqnarray}
{\cal L}_{R} &=&  -{1\over 4} (F_{\mu \nu}^{\hat a})^2   + {1 \over 2} (D_\mu \phi^{\hat a})^2  + {i } \bar \chi^p \cancel{D} \chi_p - m \bar \chi^p  \chi_p +
 g  \phi^{\hat a} \bar \chi^p t^{\hat a} \chi_p \ ,
\end{eqnarray} 
where $p= 1, \ldots , P$ is a global flavor index. The simplest case is the one in which $P=1$, which yields a non-compact gauging for the generic Jordan family theory with scalar manifold
\begin{equation}
{\mathcal{M}}_{\rm 5D}={SO(3,1) \over SO(3)} \times SO(1,1) \ .
\end{equation}
In this case, there are three relevant amplitudes between one massless and two massive vectors. The first is given by the double copy
\bea
{\cal M}_3(1\overline W,2W, 3\sugA^4)  &=&i  A_3^{{\cal N}=2}(1 \bar \chi,2  \chi, 3A) A_3^{{\cal N}=0}(1 \bar \chi,2    \chi, 3\phi)  \nn \\
&=& 
{ m}  \langle  \overline W W A \rangle_{|D W|^2}-  {1 \over 4}\langle  \overline W W A \rangle_{D\overline{W}  \wedge DW \wedge A} \  . \qquad
\eea
The result gives both a term proportional to  $F\wedge F \wedge A$  and a term proportional to the mass, which corresponds to  a minimal coupling. The second amplitude has the same expression, with the difference that  the scalar comes from the supersymmetric theory,
\bea
{\cal M}_3(1\overline W,2W, 3\sugA^1)  &=&i  A_3^{{\cal N}=2}(1  \bar \chi,2  \chi, 3 \phi) 
A_3^{{\cal N}=0}(1 \bar \chi,2  \chi, 3 A)  \nn \\
&=& -  { m}  \langle  \overline W  W A \rangle_{|D W|^2}- {1 \over 4}\langle  \overline W W A \rangle_{D\overline{W} \wedge DW \wedge A  } \ . \qquad
\eea
The third amplitude involves the $\rm 5D$ massless vector that comes from the dualization of the antisymmetric part of the double copy of two gluons,  
\bea
{\cal M}_3(1 \overline W,2W, 3\sugA^0)  &=&- i {\epsilon^{ab} \over 4} {\partial \over \partial \auxu_3^{a}} A_3^{{\cal N}=2}(1  \bar \chi,2  \chi, 3A) {\partial \over \partial \auxu_3^{b}} A_3^{{\cal N}=0}(1 \bar \chi,2  \chi, 3A) \nn \\[1pt]
&=&{ \sqrt{2}} m   \langle  \overline W W A \rangle_{W \cdot F \cdot \overline{W} } \ .
\eea
We can reproduce the amplitude in the previous section from the supergravity Lagrangian with the identification
\begin{equation}
{A}_\mu^0= {\sugA_\mu^1 + \sugA_\mu^4 \over 2}  \ ,  \qquad
{A}_\mu^1= {\sugA_\mu^1 - \sugA_\mu^4 \over 2} + {\sugA_\mu^0 \over \sqrt{2}} \ ,  \qquad
{A}_\mu^4= {\sugA_\mu^1 - \sugA_\mu^4 \over 2} - {\sugA_\mu^0 \over \sqrt{2}} \ ,
\end{equation}
provided that we rescale $\gsug = e^{-\phi} \gsug'$ and take the pp-wave limit $\phi\rightarrow \infty$ while identifying $\gsug'=-\lambda$.

 \begin{table}
\begin{center}	
\begin{tabular}{cccccc}
\bf Representation &  \bf GT 1  &\bf mass 1  & \bf  GT 2 & \bf mass 2 & \bf Supergravity \ \\
\hline
\\[-5pt]
$(N^2-1)$ & ${\cal V}_{\cN=2}$ &$0$ &$A \oplus \phi$  & $0$ & ${\cal H}_{\cN=2}\oplus2 {\cal V}_{\cN=2}$ \\
$N_1$ & ${\Phi}_{\cN=2}$ & $m_1$ & $ \chi_1$  & $-m_1$ & ${\cal V}_{\cN=2}$ \\
$\bar N_1 $ & $\bar {\Phi}_{\cN=2}$ &$-m_1$ & $\bar \chi_1$  & $m_1$ & $\overline{\cal V}_{\cN=2}$\\
\end{tabular}
\caption{Double-copy map for the alternative realization of the generic Jordan family; hypermultiplets and matter fermions are in the fundamental and antifundamental representations, respectively. Masses from the gauge theories are taken with opposite signs.\label{tab0}}	\end{center}
\end{table}

\subsection{Theories with tensors}

The simplest example of a theory with massive tensors arises from a simple modification of the construction in the previous subsection, namely flipping the sign of the mass for the fermions in the non-supersymmetric theory. The ungauged theory is the same as before and we focus again on the $P=1$ case. 

As before, the scalar manifold is 
\begin{equation}
{\cal M}_{\rm 5D}={SO(3,1) \over SO(3)} \times SO(1,1) \ .
\end{equation}
However, now we have a pair of massive tensors instead of massive vectors. Their appearance is a consequence of the gauging of a $U(1)$ subgroup of the global symmetry group of the theory under which some of the vectors are charged. The natural choice of this $U(1)$ follows from the decomposition of the isometry group as
\begin{equation}
SO(3,1) \rightarrow U(1) \times SO(1,1) \ .
\end{equation}
Two vectors are charged under $U(1)$ with opposite charges, and hence need to be dualized to tensors. Here we take $U(1)$ to be unbroken.  

The double-copy one-gluon-two-tensor amplitude is
\begin{equation}
{\cal M}_3\big( 1 \bar B,2 B,3 \sugA \big) =  i A_3^{\cN=2}\big( 1 \bar \chi ,2 \chi ,3 A \big)  A_3^{\cN=0}\big( 1 \bar \chi,2 \chi ,3 \phi \big)  =
{i \over 2}   \abra{1}\aket{2} \,  \abra{2} | \varepsilon_3 | \aket{1}  \ .
\end{equation}
As previously explained, this amplitude can be rewritten as 
\begin{equation}
{\cal M}_3\big( 1 \bar B,2 B,3 \sugA \big)  = - i m^2 \epsilon\big(\varepsilon^B_1, {\varepsilon}_3,  \varepsilon^B_2 \big)
\end{equation}
and, from the perspective of the supergravity Lagrangian, comes entirely from the $B\wedge B \wedge A$ term.  When the gauge coupling is taken to zero this amplitude becomes the three-vector amplitude in the ungauged theory.

The second amplitude has the same expression, with the difference that  the scalar comes from the supersymmetric theory.
The third amplitude involves the $\rm 5D$ vector from the  double copy of two gluons, and vanishes identically. These amplitudes are consistent with eq. (\ref{BBAamp}).

 \begin{table}
\begin{center}	
\begin{tabular}{cccccc}
\bf Representation &  \bf GT 1  &\bf mass 1  & \bf  GT 2 & \bf mass 2 & \bf Supergravity \ \\
\hline
\\[-5pt]
$(N^2-1)$ & ${\cal V}_{\cN=2}$ &$0$ &$A \oplus \phi$  & $0$ & ${\cal H}_{\cN=2}\oplus2 {\cal V}_{\cN=2}$ \\
$N_1$ & ${\Phi}_{\cN=2}$ & $m_1$ & $ \chi_1$  & $m_1$ & ${\cal T}_{\cN=2}$ \\
$\bar N_1 $ & $\bar {\Phi}_{\cN=2}$ &$-m_1$ & $\bar \chi_1$  & $-m_1$ & $\overline{\cal T}_{\cN=2}$\\
\end{tabular}
\caption{Double-copy map for the alternative realization of the generic Jordan family; hypermultiplets and matter fermions are in the fundamental and antifundamental representations; taking masses from the gauge theories with the same sign leads to tensors in the output of the double copy.\label{tabtensor}}	\end{center}
\end{table}

\subsection{Additional generalizations}

The construction outlined in the previous subsection can be generalized to include $P>1$ fermions in different representations and matching hypermultiplets of the same mass from the supersymmetric gauge theory. The fields and their masses and representations are  summarized in Table \ref{tab0b}.

It is interesting to outline how these generic masses arise from the point of view of the supergravity theory Lagrangian. The starting point is an ungauged theory with scalar manifold
\begin{equation}
{ SO(2P+1,1 ) \over SO(2P+1) } \times SO(1,1) \ ,
\end{equation}
where now we have $2P$ massive vectors and only three massless vectors. In order to have the appropriate number of masses, we need to gauge $2P+1$ generators. 
To achieve this we start from the Iwasawa decomposition $KAN$ of an element $G$ of $SO(2P+1,1)$ 
\be 
G = KAN  \ ,\qquad K\in SO(2P+1) \ ,\quad A\in  SO(1,1) \ , \quad N\in \mathbb{T}_{(2P)} \ ,
\ee
where $\mathbb{T}_{(2P)}$ is the nilpotent group generated by $2P$ translations along compact directions. 
The maximal subgroup that commutes with dilatations $SO(1,1)$ is $K_0=SO(2P)$ 
which normalizes $\mathbb{T}_{(2P)}$, {\it i.e.}
\be 
K_0 AN \ ,\qquad K_0\in SO(2P) \ ,\quad A\in  SO(1,1) \ , \quad N\in \mathbb{T}_{(2P)} \ ,
\ee
We shall gauge the subgroup of the form 
\be 
U_D(1) \ltimes \mathcal{H}_{2P+1} \ ,
\ee 
where $U_D(1)$ is a diagonal subgroup of $U(1)^P \subset SO(2P)$. 
Under the action of $U(1)^P$ the translation generators $\mathbb{T}_{(2P)}$ form $P$ doublets with charges $\pm m_i \,\,( i=2,3,4,..,P+1)$.
The resulting theory has 3 massless vector fields {\it i.e.} the gauge field of $U_D(1)$ and two massless spectators and $2P$ massive vectors.

 \begin{table}
\begin{center}	
\begin{tabular}{cccccc}
\bf Representation &  \bf GT 1 & \bf mass1  & \bf  GT 2&\bf mass 2 & \bf Supergravity  \ \\
\hline
\\[-5pt]
$(N_1^2-1)\oplus \cdots \oplus (N_P^2-1)  $  &${\cal V}_{\cN=2}$ & $0$ &$A \oplus \phi$ &$0$ & ${\cal H}_{\cN=2}\oplus 2 {\cal V}_{\cN=2}$\\
$(N_1, 1, \ldots, 1)$ & ${\Phi}_{\cN=2}$ & $m_1$ &$ \chi_1$ & $-m_1$ &  ${\cal V}^1_{\cN=2}$  \\
$(\bar N_1, 1, \ldots, 1) $ & $\bar {\Phi}_{\cN=2}$ & $-m_1$ & $\bar \chi_1$ &$m_1$ & $\overline{\cal V}^1_{\cN=2}$ \\
\vdots & \vdots & \vdots & \vdots & \vdots & \vdots\\
$(1, \ldots,1, N_P )$ & ${\Phi}_{\cN=2}$ & $m_P$ & $\chi_P$ & $-m_P$  & ${\cal V}^P_{\cN=2}$ \\
$( 1, \ldots, 1,\bar N_P) $ & $\bar {\Phi}_{\cN=2}$ & $-m_P$ & $\bar \chi_P$ &$m_P$ & $\overline{\cal V}^P_{\cN=2}$ \\
\end{tabular}
\caption{Double-copy map for the alternative realization of the generic Jordan family; we take the gauge group to be $SU(N_1)\times\dots \times SU(N_p)$ and put hypermultiplets in the fundamental representations and matter fermions in the antifundamental representations.\label{tab0b}}	\end{center}
\end{table}

Labelling the coordinate index $A$  of a vector $X_A$ of $SO(2P+1,1)$ to run from $1$ to $2P+2$ with time-like coordinate labelled as 1 and introducing complex coordinates $z_i = {1 \over \sqrt{2}} (X_{i+2} +i X_{i+P+2}) $ and $\bar{z}_i ={1 \over \sqrt{2}} (X_{i+2} -i X_{i+P+2}) $ $(i=1,\ldots,P)$
the structure constants for this gauging (obtained from three-point amplitudes) are 
\begin{eqnarray}
f^1{}_{i \bar \jmath} = - f^1{}_{\bar \jmath i} = \delta_{i \bar \jmath}  \ , \qquad  f^i{}_{2 j} = - f^i{}_{j 2} = m_i \delta^i_j  \ , 
\qquad  f^{\bar \imath}{}_{2 \bar \jmath} = - f^{\bar\imath}{}_{\bar \jmath 2} = - m_i \delta^{\bar \imath}_{\bar \jmath} \ .  
\end{eqnarray}
All other structure constants are zero. Here $1,2$ represent special (real) directions, while all other generators carry a complex index running from $1$ to $P$.\footnote{An extra coordinate $\xi^0$ is present in ambient space. We remark that the double-copy map between fields of the supergravity Lagrangian and gauge theory bilinears is different in the alternative realization of the generic Jordan family with respect to the one in Section \ref{Jordan_conventional}. In particular, here $A^0_\mu$ is not the supergravity vector obtained from the double copy of gauge-theory gluons.}
$m_i$ are free parameters (corresponding to the masses of the vectors). Jacobi relations are satisfied by this choice of structure constants. Moreover, the above gauging naturally lead to a realization with a central charge gauged by the vector field corresponding to the $\xi^1$ coordinate in ambient space. 
This construction can be straightforwardly modified to include both massive vectors and tensors.

\section{Gaugings for the complex magical theory \label{sec5}}

In this section, we move beyond the generic Jordan family, and discuss the double-copy realization of different gaugings of the complex magical theory together with their correspondence with the Lagrangian formulation of this theory.

\subsection{$U(2) \ltimes {\cal H}_5$ gauging for the complex magical theory}

For the 5D $\cN=2$ side of the double copy we use the Lagrangian (\ref{susyLhyper}),\footnote{This Lagrangian should be obtainable through a combination of Higgsing and orbifolding starting from the breaking of a larger gauge group with the procedure outlined in \rcite{Chiodaroli:2013upa}. This construction guarantees that scattering amplitudes following from this Lagrangian obey C/K duality.}
\bea
{\cal L}_{L} &=& -{1\over 4} (F_{\mu \nu}^{\hat a})^2  + i \bar \psi^{\hat a} \cancel{D} \psi^{\hat a} + {1 \over 2} (D_\mu \phi^{\hat a})^2 +  |D_\mu \varphi|^2  -  m^2 | \varphi|^2
+ {i } \bar \chi \cancel{D} \chi -  m \bar \chi \chi   \\[1pt]
&& \hskip -1.3cm +\sqrt{2} g \bar \psi^{\hat a} \varphi t^{\hat a} \chi +\sqrt{2} g \bar \chi t^{\hat a} \varphi \psi^{\hat a} + i g f^{\hat a \hat b \hat c} \phi^{\hat a} \bar \psi^{\hat b} \psi^{\hat c}  + g^2  \phi^{\hat a} \phi^{\hat b} \bar \varphi t^{\hat a} t^{\hat b} \varphi
+ g \bar \chi t^{\hat a} \chi \phi^{\hat a} + g^2 (\bar \varphi t^{\hat a} \varphi)^2  \ .
\no 
\eea
%
%
%
The moduli space of vacua is parameterized by the vevs of scalar fields:
\begin{equation}
\phi^{\hat a} \rightarrow u^{\hat a} + \phi^{\hat a}
\ ,
\qquad
\varphi \rightarrow v + \varphi \ .
\end{equation}
Minimizing the scalar potential leads to two possible solutions:
\begin{equation}
\left\{  \begin{array}{l} u^{\hat a} = 0 \\ m = 0 \end{array} \qquad \text{ or } 
\qquad v= 0 = \bar v \ .
\right.
\end{equation}
The former is the Higgs branch, the latter is the Coulomb branch, and they intersect at the origin of the moduli space, where all fields are massless.

In the following, we focus on the Coulomb branch and consider the symmetry-breaking pattern given by the adjoint-scalar vev 
\begin{equation}
\langle \phi^{\hat a} \rangle t^{\hat a} = \left( \begin{array}{cc} u_1 I_{N_1} & 0 \\ 0 & u_2 I_{N_2}  \end{array} \right)  \ .
\end{equation}
This breaks the gauge symmetry as 
\begin{equation}
U(N_1+N_2) \rightarrow SU(N_1) \times SU(N_2) \times U(1)^2 \ , \qquad
R \rightarrow (N_1,1) \oplus (1,N_2) \ ,
\end{equation} 
where the $W$-bosons transform is bifundamental representations. The masses of the fields in the various representations are given in Table \ref{tab1}. Denoting the masses of the two hypermultiplets as $m_1=m+u_1$ and $m_2=m+u_2$, we note the relation 
\begin{equation}
m_1 - m_2 = m_W \ , \label{relmasses}
\end{equation} 
which will also need to be obeyed  by the masses of the fields that double copy with the hypermultiplets.

For the non-supersymmetric side of the double copy we start from the Lagrangian
\begin{eqnarray}
{\cal L}_R &=&  
-{1\over 4} (F_{\mu \nu}^{\hat a})^2  + {1 \over 2} (D_\mu \phi^{\hat a})^2 + |D_\mu \bar \varphi|^2   - \tilde m_\varphi^2  |\varphi|^2   +  
+ {i } \bar \chi_1 \cancel{D} \chi_1 - \tilde m_1 \bar \chi_1  \chi_1 
\no \\[1pt]
&& +  
 {i } \bar \chi_2 \cancel{D} \chi_2 - \tilde m_2 \bar \chi_2  \chi_2 \no  +  a_1 g  \phi^{\hat a} \bar \chi_1 t^{\hat a} \chi_1 +  a_2 g  \phi^{\hat a}\bar \chi_2 t^{\hat a} \chi_2
+ \lambda g \phi^{\hat a} \bar \varphi T^{\hat a} \varphi   
\no \\[2pt]
&& 
+  b g   \varphi_{\hat \alpha}  \bar \chi_1  C^{\hat \alpha}  \chi_2 
+  \bar b g \bar \varphi^{\hat \alpha}  \bar \chi_2  \bar C_{\hat \alpha} \chi_1  
+ g^2 \phi^{\hat a} \phi^{\hat b}\bar \varphi T^{\hat a} T^{\hat b} \varphi  + {c \over 2} g^2 (\bar \varphi T^{\hat a} \varphi)^2 \ . 
\end{eqnarray} 
We fix the coefficients $a_1, a_2, b, {\bar b}$ and $c$ so 
that the amplitudes obey C/K duality and moreover 
the mass spectrum is the same as that of the supersymmetric side.
Here the fermions $\chi_1$ and $\chi_2$ transform in the antifundamental representations $(\bar N_1,1)$ and $(1, \bar N_2)$ respectively, for which representation matrices are denoted as $t^{\hat a}$. The massive scalars $\varphi$ are in the bifundamental representation $(\bar N_1 , N_2)$, for which the representation matrices are denotes as $T^{\hat a}$. Representations are chosen so that asymptotic states can be paired with the ones from the supersymmetric gauge theory so that supergravity states correspond to gauge-invariant bilinears. 
We find that:
\begin{itemize}
	\item The amplitude ${\cal A}_4(1 \bar \chi_i, 2  \chi_i, 3 \phi, 4 \phi  )$ has numerators
	\begin{eqnarray} \nn
	n_t &=& {1 \over 2} a_i^2 \sbra{1}| (\cancel{p}_3 + 2 \tilde m_i ) |\sket{2} \ , \\ 
	n_u &=& {1\over 2} a_i^2 \sbra{1}| (\cancel{p}_4 + 2 \tilde m_i ) |\sket{2} \ , \\
	n_s &=&  {1 \over 2} \sbra{1}| (\cancel{p}_3 - \cancel{p}_4  ) |\sket{2}\ . \nn
	\end{eqnarray}
 The kinematic Jacobi relation $n_t-n_u=n_s$ implies $a_i=\pm 1$. Note that leg one is taken with a mass of $-m_i$ and leg two is taken with a mass of $+m_i$ so that masses add to zero. 
	\item The amplitude ${\cal A}_4(1 \bar \chi_1, 2  \chi_1, 3  \bar \varphi, 4  \varphi  )$ has numerators
	\begin{eqnarray}
	n_t &=& {1 \over 2} |b|^2 \sbra{1}| (\cancel{p}_3 + \tilde m_1+\tilde m_2 )|\sket{2} \ , \nn \\ 
	n_s &=& {1 \over 2} \sbra{1}| (2\cancel{p}_3 + a_1 \lambda  )|\sket{2} \ .
	\end{eqnarray}
	C/K duality $n_t=n_s$ implies $|b|^2=2$ and the condition
	\begin{equation}
	\tilde m_1 + \tilde m_2 = {\lambda \over 2} \ . \label{lambdacond}
	\end{equation} 
		\item The amplitude ${\cal A}_4(1 \bar \chi_1, 2  \chi_2, 3 \varphi, 4 \phi  )$ has numerators
	\begin{eqnarray}
	n_t &=& {1\over 2} a_1  b \ \sbra{1}| (\cancel{p}_3 +  \tilde m_1  + \tilde m_2 ) |\sket{2} \ , \nn \\ 
	n_u &=& {1 \over 2}a_2   b \ \sbra{1}| (\cancel{p}_4 + 2 \tilde m_2 )  |\sket{2} \ , \\ 
	n_s &=& {1 \over 2} b \lambda \ \sbra{1}\sket{2}  \ . \nn
	\end{eqnarray}
	C/K duality $n_t-n_u=n_s$ implies the additional condition $a_1=-a_2$. 
\item The amplitude ${\cal A}_4(1 \bar \varphi, 2  \bar \varphi, 3 \varphi, 4 \varphi  )$ has numerators ($i=1,2$)
\begin{eqnarray}
n_t &=& {1\over 2} (s - u + \lambda^2 + c\, t )\ , \nn \\ 
n_u &=&  {1 \over 2} (s - t + \lambda^2 + c\, u) \ . 
\end{eqnarray}
C/K duality $n_t=n_u$ implies the condition $c=-1$. 
\end{itemize}
Note that the mass $\tilde m_\varphi$ of the bifundamental scalar does not obey any 
constraint other than the one following from matching the mass spectra of the two
theories. 
We can rewrite the amplitudes we have analyzed above in the $\tilde m_1=\tilde m_2$ case in a form that is explicitly covariant under the unbroken $SU(2)$ symmetry by expressing the complex scalar fields in terms of real fields as $\varphi={1 \over \sqrt{2}} (\phi^1 - i \phi^2 )$ and identifying $\phi^3=\phi$. The resulting amplitude ${\cal A}_4(1 \bar \chi, 2  \chi, 3 \phi^i, 4 \phi^j  )$ has numerators 
	\begin{eqnarray}
n_t &=&   {1 \over 2} \sbra{1}| (\cancel{p}_3 +  2 \tilde m ) |\sket{2} \ {\rm t}^j {\rm t}^i \ , \label{numcovariantt} \\ 
n_u &=&   {1\over 2} \sbra{1}| (\cancel{p}_4 + 2 \tilde m )  |\sket{2} \ {\rm t}^i {\rm t}^j \ ,  \\ 
n_s &=&    \sbra{1}|  \cancel{p}_3 |\sket{2}  \delta^{ij} - {i \over 2} \lambda  \sbra{1} \sket{2} \epsilon^{ijk} {\rm t}^k   \ , \label{numcovariants}
\end{eqnarray}
where ${\rm t}^i=\sigma^i$ are $SU(2)$ representation matrices.

The double copy combines fields with in conjugate representations of the gauge group
and with equal masses up to a sign. The resulting supergravity fields and their gauge-theory origin are collected in Table~\ref{tab1}. We now focus on three-point amplitudes containing vectors obtained from the double copy and consider the case in which $u_1=u_2$. A first amplitude comes from the bosonic double copy,
\begin{equation}
{\cal M}_3(1 \sugA^{i+1}\!\!,2 \sugA^{j+1}\!\!,3 {\sugA^{k+1}}) \!= \!
i A_3^{{\cal N}=2}(1 A,2  A, 3A) A_3^{{\cal N}=0}(1 \phi^i \!\!,2    \phi^j\!\!, 3\phi^k)   = - {\lambda \over 2} \epsilon^{ijk} \langle \overline W W A \rangle_{\rm YM} 
\end{equation}
where all vectors and scalar fields are now taken to be massless and  $i,j,k=1,2,3$. Fermionic double copies give amplitudes of the form 
\bea
{\cal M}_3(1\overline W,2W, 3\sugA^{i+1})  &=&i  A_3^{{\cal N}=2}(1 \bar \chi,2  \chi, 3A) A_3^{{\cal N}=0}(1 \bar \chi,2    \chi, 3\phi^i) \no \\ &=&
{ m} \sigma^i \langle  \overline W W A \rangle_{|D W|^2}-  { \sigma^i \over 4}\langle  \overline W W A \rangle_{D\overline{W}  \wedge DW \wedge A}  , \no\\   \qquad
{\cal M}_3(1\overline W,2W, 3\sugA^1)  &=&i  A_3^{{\cal N}=2}(1  \bar \chi,2  \chi, 3 \phi) 
A_3^{{\cal N}=0}(1 \bar \chi,2  \chi, 3 A)  \nn \\
&=& -  { m}  \langle  \overline W  W A \rangle_{|D W|^2}- {1 \over 4}\langle  \overline W W A \rangle_{D\overline{W} \wedge DW \wedge A  }  , \no \\ 
{\cal M}_3(1 \overline W,2W, 3\sugA^0)  &=&- i {\epsilon^{ab} \over 4} {\partial \over \partial \auxu_3^{a}} A_3^{{\cal N}=2}(1  \bar \chi,2  \chi, 3A) {\partial \over \partial \auxu_3^{b}} A_3^{{\cal N}=0}(1 \bar \chi,2  \chi, 3A) \nn \\
&=&{ \sqrt{2}} m   \langle  \overline W W A \rangle_{W \cdot F \cdot \overline{W} } \ . \label{ampsdccomplex}
\eea
We shall see how the same amplitudes can be obtained from the supergravity Lagrangian.

\begin{table}
	\begin{center}	
 \
 
		\begin{tabular}{@{}c@{}ccccc@{}}
		\textbf{Representation} & \bf \ GT 1   &  \bf mass 1 &  \bf  GT 2  \ & \bf mass 2 & \bf Supergravity \ \\
			\hline
			\\[-5pt]
			Adj. & ${\cal V}_{\cN=2}$ & $0$ & $A_\mu \oplus \phi$  & $0$ & ${\cal H}_{\cN=2} \oplus 2{\cal V}_{\cN=2}$ \\		
			$(N_1, \bar N_2)$ & ${\cal V}^{(m)}_{\cN=2}$ & $m_W=u_1 - u_2$ & $\varphi$ & $\tilde m_\varphi=- m_W$ & ${\cal V}^{(m)}_{\cN=2}$ \\
			$(\bar N_1, N_2)$ & $\overline{\cal V}^{(m)}_{\cN=2}$ & $-m_W$ & $\bar \varphi$ & $- \tilde m_\varphi$ & $\overline{\cal V}^{(m)}_{\cN=2}$ \\
			$(N_1, 1 )$ & ${\Phi}_{\cN=2}$ &$m_1=m + u_1$ & $\chi_1$ &$\tilde m_1=-m_1$ & ${\cal V}^{(m)}_{\cN=2}$ \\
			$(\bar N_1, 1 )$ & $\bar {\Phi}_{\cN=2}$& $- m_1$ & $\bar\chi_1$ & $-\tilde m_1$ & $\overline{\cal V}^{(m)}_{\cN=2}$\\
			$(1, N_2 )$ & ${\Phi}_{\cN=2}$ &$m_2=m + u_2$ & $\chi_2$ & $\tilde m_2 = -m_2$ &
			${\cal V}^{(m)}_{\cN=2}$\\
			$(1, \bar N_2 )$ & $\bar {\Phi}_{\cN=2}$ &$-m_2$ &$\bar \chi_2$ & $-\tilde m_2$ & $\overline{\cal V}^{(m)}_{\cN=2}$ \\
		\end{tabular}
		\caption{Double-copy construction for spontaneously broken $U(2) \ltimes {\cal H}_5 \rightarrow U(1)^2\ltimes {\cal H}_3$ gauging of the complex magical theory. Masses in the two gauge theories are matched with opposite signs.
  Fields in GT 1 are taken in the representations listed in the table, while GT 2 fields are taken in the conjugate representations. 
  The Adj. representation is $(N_1^2-1,1) \oplus	 (1,N_2^2-1) \oplus (1,1)$. \label{tab1} The supergravity gauge group $U(2)$ is unbroken when $u_1=u_2$.}	
	\end{center}
\end{table}

\subsection{Supergravity identification}

The starting point is to write the cubic polynomial with an explicit $U(2,1)$ symmetry. 
To this end, we introduce the matrix 
\begin{equation}
N(\xi) = \left( \begin{array}{ccc} 
\xi^1 - \xi^4 & -\xi^2 + i \xi^3 &  -\xi^5 \\
-\xi^2 - i \xi^3 &  \xi^1 + \xi^4 & - \xi^6 \\
\bar \xi^5 &  \bar \xi^6  &  \sqrt{2}\xi^0 
\end{array}  \right) \ .
\end{equation}
The cubic polynomial in the canonical basis is then given by 
\begin{equation}
{\cal V} (\xi) = \det{N(\xi)} \ ,
\end{equation}
while a $U(2,1)$ transformation $V$ acts on $N$ as
\begin{equation}
 N(\xi) \rightarrow V^{-1} N(\xi) V  \ ,
\end{equation}
The  $U(2,1)$ generators associated to each direction in ambient space are then given by\footnote{This expression can be obtained from the conventional organization of the ambient-space coordinates as a Hermitian $3 \times 3$ matrix by a field redefinition. Making manifest the $U(2,1)$ action comes at the price of obfuscating the connection with Euclidean Jordan algebras from \rcite{Gunaydin:1983bi}.} 
\begin{eqnarray}
T_I = -{i \over 2} {\partial {N(\xi)} \over \partial \xi^I }   \ .
\end{eqnarray} 

We choose a base-point that depends on 2 parameters $\theta, \phi$, such that 
\begin{equation}
N(c^I) =  \text{diag}\big( -e^{\theta/3 + \phi}, \ -e^{\theta/3 - \phi} , \  e^{-2\theta/3}  \big) \ . 
\end{equation}
Repeating the steps described in Section~\ref{sec5dMESGT} we obtain expressions for $\å_{IJ}$ and $K_I^x$. It is convenient to introduce a linear field transformation that brings the vector-field kinetic term to the canonical form, $A_\mu \rightarrow U {A}_\mu$,
\begin{equation}
U= \left(
\begin{array}{c@{}c@{}c@{}c@{}c@{}c@{}c@{}c@{}c}
{e^{- {2\over 3} \theta} } & 0  & 0 & 0 & 0  & 0 & 0 & 0 & 0 \\
0 & e^{\theta \over 3} \cosh{\phi} & 0 & 0 & - e^{\theta \over 3} \sinh{\phi} & 0 & 0 & 0 & 0 \\
0 & 0 &  e^{\theta \over 3} & 0 & 0 & 0 & 0 & 0 & 0 \\
0 & 0 & 0 & 	e^{\theta \over 3} & 0 & 0 & 0 & 0 & 0 \\
0 & - e^{\theta \over 3} \sinh{\phi} & 0 & 0 &  e^{\theta \over 3} \cosh{\phi} & 0 & 0 & 0 & 0 \\ 
0 & 0 & 0 & 0 & 0 & \sqrt{2} e^{-{\theta \over 6} +{\phi \over 2}}  & 0 & 0 & 0 \\
0 & 0 & 0 & 0 & 0 & 0 & 	\sqrt{2} e^{-{\theta \over 6} - {\phi \over 2}} & 0 & 0 \\
0 & 0 & 0 & 0 & 0 & 0 & 0 & \sqrt{2} e^{-{\theta \over 6} + {\phi \over 2}}  & 0 \\ 
0 & 0 & 0 & 0 & 0 & 0 & 0 & 0 & \sqrt{2} e^{-{\theta \over 6} - {\phi \over 2}}  
\end{array} \right) \ . \no
\end{equation}
This further requires a rescaling of the supergravity gauge coupling
\begin{equation}
\gsug = e^{-{2\theta \over 3}} \gsug' \ .
\end{equation}
After these transformations, the mass matrix for the supergravity vector fields has mostly vanishing entries except for 
\begin{equation}
m_{22} = m_{33} =   \gsug'  \sinh{\phi} \ , \quad
m_{5\bar 5} =  \gsug' e^{\phi - \theta \over 2} \cosh {{\theta + \phi \over 2}} \ ,\quad
m_{6 \bar 6} =  \gsug' e^{-{\phi + \theta \over 2}} \cosh {{\theta - \phi \over 2}} \ .
\end{equation}
We note that the parameter $\phi$ is related to the spontaneous breaking of the compact subgroup $U(2)$; the unbroken phase is obtained with $\phi=0$. The three masses are not independent. They obey the simple relation
\begin{equation}
m_{5 \bar 5} - m_{6 \bar 6 } = m_{22} \ ,
\end{equation}
which we have seen emerge naturally from the double-copy framework in eq. (\ref{relmasses}).

The three-point amplitudes between the three $SU(2)$ fields are
\begin{equation}
{\cal M}_3(1{A^{i+1}},2 {A^{j+1}},3 {A^{k+1}}) = i  \gsug'   \cosh \phi \  \epsilon^{ijk}  \langle \overline W W A \rangle_{\rm YM} \label{3vec} \ ,
\end{equation}
with $i,j,k=1,2,3$. 
Additionally, we have four massive vectors 
that transform in the fundamental of $SU(2)$ and the anti-fundamental. We denote them as $W$ and $\overline{W}$ and do not display the $SU(2)$ indices explicitly. The extra amplitudes that involve the massive vectors are given in the $\phi=0$ case for notational convenience, 
\begin{eqnarray}
{\cal M}_3(1{\overline{W}},2{ W},3{A}^{i+1}) &=&  {1 \over 2}\gsug' \sigma^{i} \Big\{ e^{-{\theta }}\langle \overline W W A \rangle \Big|_{W \cdot F \cdot \overline{W} } - \langle \overline W  W A \rangle \Big|_{|D W|^2} \Big\}\ , \no \\
{\cal M}_3(1{\overline{W}},2{ W},3{A}^0) &=&  {1  \over \sqrt{2}}\gsug' \mathbb{I}_2 \Big\{    e^{-{\theta}} \langle \overline W  W A \rangle \Big|_{|D W|^2}  -  \langle \overline W W A \rangle \Big|_{W \cdot F \cdot \overline{W} }  \Big\}\ ,  \nn\\
{\cal M}_3(1{\overline{W}'},2{ W'},3{A'}^1) &=& - {1 \over 2}\gsug' \mathbb{I}_2 \Big\{     e^{-{\theta}} \langle \overline W W A \rangle \Big|_{W \cdot F \cdot \overline{W} } - \langle \overline W  W A \rangle \Big|_{|D W|^2} \Big\}\ . 
\end{eqnarray}
Amplitudes can be matched with the double-copy formulae (\ref{ampsdccomplex}) only in the $\theta\rightarrow \infty$ limit. 
The masses are then matched provided that
\begin{equation}
2m +{ u_1 + u_2  } = -{\lambda \over 2} = {\gsug' }  \cosh{\phi}   \ ,\qquad {u_1 - u_2 }= {\gsug' } \sinh \phi  \ .
\end{equation}
Note that the difference between $U(2,1)$ and this contraction is detectable only at the level of three-point amplitudes and cannot be seen from the spectra alone.

\subsection{Non-compact gaugings with tensors}

The simplest example of gauging which interchanges massive non-compact gauge fields with tensors is a $U(2)$ gauging of the complex magical theory; this gauging is obtained by flipping the signs for the fermionic masses in the non-supersymmetric gauge-theory factor entering the double copy. The resulting spectrum can be found in Table \ref{tab6a}.

\begin{table}
	\begin{center}	
		\begin{tabular}{@{}c@{}ccccc@{}}
			\bf Representation & \bf \ GT 1   &  \bf mass 1 &  \bf  GT 2  \ & \bf mass 2 & \bf Supergravity \ \\
			\hline
			\\[-5pt]
		Adj. 	& ${\cal V}_{\cN=2}$ & $0$ & $A_\mu \oplus \phi$  & $0$ & ${\cal H}_{\cN=2} \oplus 2{\cal V}_{\cN=2}$ \\	
			$(N_1, \bar N_2)$ & ${\cal V}^{(m)}_{\cN=2}$ & $m_W=u_1 - u_2$ & $\varphi$ & $\tilde m_\varphi=-m_W$ & ${\cal V}^{(m)}_{\cN=2}$ \\
			$(\bar N_1, N_2)$ & $\overline{\cal V}^{(m)}_{\cN=2}$ & $-m_W$ & $\bar \varphi$ & $- \tilde m_\varphi$ & $\overline{\cal V}^{(m)}_{\cN=2}$ \\
			$(N_1, 1 )$ & ${\Phi}_{\cN=2}$ &$m_1=m + u_1$ & $\chi_1$ &$\tilde m_1 = m_1$ & ${\cal T}^{(m)}_{\cN=2}$ \\
			$(\bar N_1, 1 )$ & $\bar {\Phi}_{\cN=2}$& $-m_1$ & $\bar\chi_1$ & $-\tilde m_1$ & $\overline{\cal T}^{(m)}_{\cN=2}$\\
			$(1, N_2 )$ & ${\Phi}_{\cN=2}$ &$m_2=m + u_2$ & $\chi_2$ & $\tilde m_2 = m_2$ &
			${\cal T}^{(m)}_{\cN=2}$\\
			$(1, \bar N_2 )$ & $\bar {\Phi}_{\cN=2}$ &$-m_2$ &$\bar \chi_2$ & $- \tilde m_2$ & $\overline{\cal T}^{(m)}_{\cN=2}$ \\
		\end{tabular}
		\caption{Double-copy construction for the spontaneously broken $U(2) \rightarrow U(1)^2$ gauging of the complex magical theory with tensors. The theory is unbroken when $u_1=u_2$. \label{tab6a}}	
	\end{center}
\end{table}

Note that this theory has a $U(2)$ compact gauge symmetry that can generically be spontaneously broken. Four tensor fields transform in the ${\bf 2}$
and $\bar{\bf 2}$ representations.
To identify the corresponding supergravity Lagrangian we write the cubic polynomial as the determinant of the matrix 
\begin{equation}
N(\xi) = \left( \begin{array}{ccc} 
\xi^1 + \xi^4 & \xi^2 - i \xi^3 &  \xi^5 \\
\xi^2 + i \xi^3 &  \xi^1 - \xi^4 &  \xi^6 \\
\bar \xi^5 &  \bar \xi^6  &  \sqrt{2}\xi^0 
\end{array}  \right)
\end{equation}
and
take the following base-point $c^I$ in the ambient space\footnote{We emphasize that we have not chosen the most general base-point.}
\begin{equation}
N(c) =  \text{diag}\big(  e^{{\theta \over 3} + \phi}, \  e^{{\theta \over 3} - \phi} , \  e^{-{2\over 3}\theta}  \big) \ .
\end{equation}
The steps taken in finding the 
 $\å_{IJ}$ and $K_I^x$ are the same as in the previous section, with the difference that only a compact $U(2)$, with gluons corresponding to the $\xi^1, \ldots, \xi^4$ coordinates in ambient space, is gauged. The fields corresponding to the coordinates $\xi^5, \ldots, \xi^8$ are now dualized to tensors, with the entries of the matrix $\å_{IJ}$ giving the values for the corresponding masses. As before, 
 a linear field redefinition is needed to bring 
 the kinetic term for the remaining vector fields to the canonical form, $A_\mu  U {A}_\mu$,
\begin{equation}
U= \left(\rightarrow
\begin{array}{ccccc}
{e^{ - {2 \over 3 } \theta} } & 0  & 0 & 0 & 0   \\
0 & e^{\theta \over 3} \cosh{\phi} & 0 & 0 & e^{\theta \over 3 } \sinh{\phi} \\
0 & 0 &  e^{\theta \over 3} & 0 & 0  \\
0 & 0 & 0 & 	e^{\theta \over 3} & 0  \\
0 &  e^{\theta \over 3} \sinh{\phi} & 0 & 0 &  e^{\theta \over 3 } \cosh{\phi}  \\ 
\end{array} \right) \no \ .
\end{equation}
After these operation, the mass matrix for the massive $W$ vectors becomes as before
\begin{equation}
m_{22} = m_{33} =   \gsug  e^{{\theta \over 3}}\sinh{\phi} \ ,\end{equation}
where all other entries are zero. 
Similarly,  $\phi$ is related to the spontaneous breaking of the compact group $U(2)$. 
The masses for the tensors can be obtained from the  corresponding entries of the $\å_{IJ}$ matrix; they are
\begin{equation}
m_{T \pm} = {\gsug \over 2} e^{{\theta \over 3} \pm \phi} \ .
\end{equation}
Again, the three masses are related by Eq.~(\ref{relmasses}) which is inherited from the spontaneously broken gauge theory entering the double-copy construction. In this case, it is not necessary to take a particular limit to obtain the supergravity Lagrangian which matches the output from the double copy. Instead, the parameter $\theta$ can be absorbed by a redefinition of $\gsug$.   The presence of a $U(1)$ factor in the gauge group 
cannot be avoided, as pointed out in \rcite{Ellis:2001xd}. From the perspective of the double-copy construction this is a consequence of the presence of a scalar in the supersymmetric theory, which produces a vector in the double-copy theory that is coupled to the massive tensors.

\subsection{$U(1,1)\ltimes {\cal H}_3$ gauging of the complex theory with tensors \label{sec-hybrid}}

\begin{table}
	\begin{center}	
		\begin{tabular}{@{}c@{}ccccc@{}}
			\bf Representation & \bf \ GT 1   &  \bf mass 1 &  \bf  GT 2  \ & \bf mass 2 & \bf Supergravity \ \\
			\hline
			\\[-5pt]
			Adj. & ${\cal V}_{\cN=2}$ & $0$ & $A_\mu \oplus \phi$  & $0$ & ${\cal H}_{\cN=2} \oplus 2{\cal V}_{\cN=2}$ \\	
			$(N_1, \bar N_2)$ & ${\cal V}^{(m)}_{\cN=2}$ & $m_W=u_1 - u_2$ & $\varphi$ & $\tilde m_\varphi = -m_W$ & ${\cal V}^{(m)}_{\cN=2}$ \\
			$(\bar N_1, N_2)$ & $\overline{\cal V}^{(m)}_{\cN=2}$ & $- m_W$ & $\bar \varphi$ & $-\tilde m_\varphi$ & $\overline{\cal V}^{(m)}_{\cN=2}$ \\
			$(N_1, 1 )$ & ${\Phi}_{\cN=2}$ &$m_1=m + u_1$ & $\chi_1$ &$\tilde m_1=-m_1$ & ${\cal V}^{(m)}_{\cN=2}$ \\
			$(\bar N_1, 1 )$ & $\bar {\Phi}_{\cN=2}$& $-m_1$ & $\bar\chi_1$ & $-\tilde m_1$ & $\overline{\cal V}^{(m)}_{\cN=2}$\\
			$(1, N_2 )$ & ${\Phi}_{\cN=2}$ &$m_2 =m + u_2$ & $\chi_2$ & $\tilde m_2=m_2$ &
			${\cal T}^{(m)}_{\cN=2}$\\
			$(1, \bar N_2 )$ & $\bar {\Phi}_{\cN=2}$ &$-m_2$ &$\bar \chi_2$ & $-\tilde m_2$ & $\overline{\cal T}^{(m)}_{\cN=2}$ \\
		\end{tabular}
		\caption{Double-copy construction for $U(1,1) \ltimes {\cal H}_3$ gauging with tensors of the complex magical theory. Here the Adj. representation is $(N_1^2-1,1) \oplus  (1,N_2^2-1) \oplus (1,1)$. \label{tab1c}}	
	\end{center}
\end{table}

We have seen that the choice of signs in matching masses between the two gauge theories entering the double copy can have significant consequences. Considering all possible ways of doing so, one may wish to study a third option in which an equal number of massive tensors and vectors is realized, in the double-copy sense, as direct product of fermions. This is done by matching one of the masses with the same signs, and the other with opposite signs, as illustrated in Table \ref{tab1c}. This choice changes the constraints imposed by C/K duality, which now become
\begin{equation}
\lambda = \tilde m_1 + \tilde m_2 = u_2 -u_1 \ .
\end{equation}
How can we characterize this new double-copy theory?
Unlike previous constructions, setting $u_1 = u_2$ forces the coupling between three gluons to vanish. Hence, the theory does not have an unbroken $U(2)$ phase. Instead, the gauge group is the non-compact 
$U(1,1)$ and three gluons become spectators when $u_1=u_2$. For all values of the parameters, this gauging has two massive vectors and a central charge vector gauging the Heisenberg group ${\cal H}_3$.  We also note that truncating away all fermionic fields gives the non-compact $SO(2,1)$ gauging of the generic Jordan family, whereas the constructions in the previous subsections would give compact $SO(3)$ gaugings.

A striking feature of this hybrid construction is that it gives the nonvanishing amplitude
\begin{equation}
{\cal M}_3 (1 \bar B, 2 W, 3W) = {i\over 2} \abra{1}\aket{2} \sbra{2} |  { \varepsilon}_3 | \aket{1} \ ,
\end{equation}
which corresponds to a coupling between two vectors and a tensor. The third vector is in general massive, but can be made massless by choosing $u_1=u_2$. This is an example of exotic vector-vector-tensor couplings introduced in \rcite{Bergshoeff:2004kh}.

\section{Gaugings for the quaternionic magical theory \label{sec6}}

We turn our attention to gaugings for the quaternionic magical theory. As usual we are constrained by the fact that there is a limit to the rank of the unbroken gauge theory coming from the double-copy construction which is in turn linked to the number of scalar fields of the non-supersymmetric gauge-theory factor, {\it i.e.} the non-compact gauge group can be at most $U(2,1)$.

The supersymmetric gauge theory is the same we have utilized for the complex theory. For the non-supersymmetric gauge-theory factor, we take a YM+fermion theory in 10D and reduce it to five dimensions. The five dimensional fields are
\begin{equation}
\big\{ A_\mu , \ \phi^i , \ \chi_{\alpha r a } \big\} \ ,
\end{equation}
where $i=1,\ldots,5$ is an index running over the compact directions, $\alpha$ is a 5D spacetime spinor index, $a=1,\ldots, 4$ is a spinor index corresponding to the five compact directions, $r=1,2$ is a an extra index which is eventually fixed by imposing the 10D chirality condition. Details about gamma matrices can be found in Appendix~\ref{appB}.  
The Lagrangian has the form 
\begin{eqnarray}
{\cal L}_{R} &=&  
-{1\over 4} (F_{\mu \nu}^{\hat a})^2  + {1 \over 2} (D_\mu \phi^{\hat a i})^2    
+ {i } \bar \chi \cancel{D} \chi - \bar \chi \tilde M  \chi 
-  g  \phi^{\hat a i} \bar \chi t^{\hat a} \tilde \gamma^i \chi  \no \\
&& + 
 \lambda g \tilde F^{i+4,j+4,k+4} f^{\hat a \hat b \hat c} \phi^{\hat a i } \phi^{\hat b j} \phi^{\hat c k} + {\cal O} (\phi^4) \ . 
\end{eqnarray} 
where $\tilde \gamma$ and $\tilde M$ are matrices acting on the internal $a,b$ spinor indices.  They can be obtained from the $D$-dimensional gamma matrices $\Gamma^I$ and mass matrix $M=- i \mathbb{I}\otimes \sigma^2 \otimes \tilde M$ obeying the 
C/K duality condition~\eqref{masterintro}, as explained in Appendix \ref{appB}.
$F^{IJK}$ is the tensor giving the symmetric trilinear scalar couplings. We choose $F^{IJK}$ and $u^I$ in eq. \eqref{fermionmass}  so that the only non-zero entries are $F^{789}=-1$ and $u^9=u$. The result is
\begin{equation}
\tilde M =  {\lambda \over 4} ( \sigma^3  \otimes 1) + u ( \sigma^3  \otimes \sigma^3 )  \ ,
\end{equation}
where our choice of gamma matrices, reported in Appendix \ref{appB}, is such that the mass matrix is diagonal. In contrast to the complex case, our spinors have now two indices, the first of which corresponds to a $SU(2)$ group that will be gauged in the output of the double copy. An important property of the mass matrix above is that it is traceless, so that mass eigenvalues come in pairs with opposite signs. 
Thus, the spectrum of the double-copy theory contains both vectors and tensors.

The next step is to orbifold the theory with respect to the group element acting on the fields of the theory as
\begin{equation}
 \Phi \rightarrow R^\Phi_{34}(\theta) g^\dagger \Phi g \ , \qquad 
 \Psi \rightarrow R^\Psi_{34}(\theta) g \Psi \ , \qquad 
 g=\text{diag} \big( e^{i \theta} I_N , e^{- i \theta} I_N \big) \ ,
\end{equation}
where $\Phi$ is a generic adjoint field, $\Psi$ is a generic fundamental field, 
$R^\Phi_{34}(\theta)$ and $R^\Psi_{34}(\theta)$ are rotations by some angle $\theta$ 
in the $3-4$ plane in the compact directions in the corresponding representations of the fields $\Phi$ and $\Psi$ of the rotation group, and $g$ is the realization of that rotation as a subgroup of the gauge group. We may choose $\theta = 2 \pi/n$ for any $n\ge 3$.
This operation breaks the gauge group as 
\begin{equation}
 SU(2N) \rightarrow SU(N) \times SU(N) \times U(1) \ .
\end{equation}
Since they carry nontrivial $R_{12}$ charge, the surviving scalars $\phi^{3,4}$ are in bifundamental representations and the surviving fermions are in fundamental representations. It is possible to add a mass term $\tilde m_\varphi^2 {\rm Tr} [\bar \varphi \varphi ]$ for the complex scalar built as $\varphi = {1 \over \sqrt{2}} (\phi^3 - i \phi^4) $.
This deformation corresponds to spontaneously breaking the supergravity gauge group. Matching the masses requires the identification
\begin{equation}
m + {u_1 +u_2 \over 2} = - {\lambda \over 4} \ , \qquad {u_1 - u_2 \over 2} = - \tilde u \ 
\end{equation}   
between the parameters in the left and right gauge theory.  The output of the construction is summarized in Table \ref{tab2}. For $\tilde m_\varphi = 0 $ we have a theory in which the gauge group is $U(2) \ltimes \mathcal{H}_{5}$.

\begin{table}
	\begin{center}	
		\begin{tabular}{@{}ccc@{}ccc@{}c@{}}
			\bf Representation & \bf   GT1  &  \bf mass 1 &  \bf   GT2  &  \bf mass 2 & \bf Supergravity   \ \ \\
			\hline
			\\[-5pt]
			Adj. & ${\cal V}_{\cN=2}$ & $0$ & $A_\mu \oplus \phi^5 \oplus $ & $0$ & ${\cal H}_{\cN=2} \oplus 4{\cal V}_{\cN=2}$  \\
			 & & & $z \oplus \bar z$ &  & & \\
			$(N_1, \bar N_2)$ & ${\cal V}^{(m)}_{\cN=2}$  & $m_W=u_1-u_2$ & $ \varphi$ & $\tilde m_\varphi=-m_W$ & ${\cal V}^{(m)}_{\cN=2}$ & \\
			$(\bar N_1, N_2)$ & $\overline{\cal V}^{(m)}_{\cN=2}$  & $-m_W$ & $\bar \varphi$ & $-\tilde m_\varphi$ & $\overline{\cal V}^{(m)}_{\cN=2}$ & \\
			$(N_1, 1 )$ & ${\Phi}^{(m)}_{\cN=2}$ & $m_1 = m +u_1$ & $\chi_{11}$ & ${\lambda \over 4} + \tilde u = - m_1$ & ${\cal V}^{(m)}_{\cN=2}$ \\
			$(N_1, 1 )$ & ${\Phi}^{(m)}_{\cN=2}$ & $m_1$ & $ \chi_{21} $& $-{\lambda \over 4} - \tilde u = m_1$ & ${\cal T}^{(m)}_{\cN=2}$\\
			$(\bar N_1, 1 )$ & $\overline{\Phi}^{(m)}_{\cN=2}$ & $-m_1$ & $\bar \chi^{11}$ & $-{\lambda \over 4} - \tilde u$ & $\overline{\cal V}^{(m)}_{\cN=2}$ \\
			$(\bar N_1, 1 )$ & $\overline{\Phi}^{(m)}_{\cN=2}$ & $-m_1$ &  $\bar \chi^{21}$ & ${\lambda \over 4} + \tilde u$ & $\overline{\cal T}^{(m)}_{\cN=2}$ \\
			$(1, N_2 )$ & ${\Phi}^{(m)}_{\cN=2}$& $m_2=m+u_2$ &  $ \chi_{12}$ & ${\lambda \over 4} -\tilde u = -m_2$ & ${\cal V}^{(m)}_{\cN=2}$  \\
			$(1, N_2 )$ & ${\Phi}^{(m)}_{\cN=2}$& $m_2$ & $ \chi_{22}$ & $-{\lambda \over 4} + \tilde u = m_2$ & ${\cal T}^{(m)}_{\cN=2}$ \\
			$(1, \bar N_2 )$ &$\overline{\Phi}^{(m)}_{\cN=2}$ & $-m_2$ & $\bar \chi^{12}$ & $-{\lambda \over 4} + \tilde u$ & $\overline{\cal V}^{(m)}_{\cN=2}$ \\
			$(1, \bar N_2 )$ &$\overline{\Phi}^{(m)}_{\cN=2}$& $-m_2$ &  $\bar \chi^{22}$ & ${\lambda \over 4} - \tilde u$ & $\overline{\cal T}^{(m)}_{\cN=2}$ \\
		\end{tabular}
		\caption{Double-copy construction for the $U(2)$ gauging of the quaternionic theory with tensors.\label{tab2} For $m_\varphi\neq0$ $U(2)$ is spontaneously broken to $U(1)^2$. The Adj. representation is $(N_1^2-1,1) \oplus  (1,N_2^2-1) \oplus (1,1)$. \label{tab3}}
	\end{center}

\end{table}

The two constructions outlined in the previous section can be obtained as truncations removing massive vectors or tensors of the one described here. In this respect, it is straightforward to identify the base-point for the supergravity theory using the matching we have already worked out.

To work out this example more explicitly, we take $u_1=u_2$ and computing amplitudes involving vectors and tensors at three points. The amplitude between three gauge bosons is
\begin{eqnarray}
{\cal M}_3\big( 1 \sugA^{i+3},2 \sugA^{j+3},3 \sugA^{k+3} \big) && = i A_3^{\cN=2}\big( 1 A,2 A,3 A \big)   A_3^{\cN=0}\big( 1 \phi^{i+2},2 \phi^{j+2},3 \phi^{k+2} \big)  \no \\ 
&&=  -{\lambda \over 2} \epsilon^{ijk}  \langle \overline W W A \rangle_{\rm YM} \ .
\end{eqnarray}
Then, the amplitude between one gluon and two massive vectors is
\begin{equation}
{\cal M}_3\big( 1 \overline{W},2 W,3 \sugA^{j+3} \big) =i  A_3^{\cN=2}\big( 1 \bar \chi ,2 \chi ,3 A \big)  A_3^{\cN=0}\big( 1 \bar \chi,2 \chi ,3 \phi^{j+2} \big)  =  {i \over 2} \sigma^j \sbra{1}\sket{2} \,  \abra{2} | \varepsilon_3 | \aket{1} \ .
\end{equation}
Here $W$ and $\overline{W}$ represents the $SU(2)$ doublets $(W_{1},W_{2})$ and $(\overline{W}_{1}, \overline{W}_2)$ and $\sigma^i$ are Pauli matrices acting on the $SU(2)$ indices. We note the relation 
\begin{equation}
t^i = \{ \tilde \gamma^i, \tilde M \} \ ,
\end{equation}
connecting the Dirac matrices in higher-dimension, the fermion mass matrix and the representation matrices of the supergravity gauge group. 
This notation makes it explicit that the supergravity $SU(2) \subset U(2,1)$ gauge group is unbroken for the choice $u_1=u_2$. 
We also have
\begin{eqnarray}
{\cal M}_3\big( 1 \overline{W},2 W,3 \sugA^1 \big) = i A_3^{\cN=2}\big( 1 \bar \chi ,2 \chi ,3 \phi \big)  A_3^{\cN=0}\big( 1 \bar \chi,2 \chi ,3 A \big) ={i \over 2}  \abra{1}\aket{2} \,  \sbra{1} | \varepsilon_3 | \sket{2} \ .\no \\
\end{eqnarray}
This amplitude indicates that the unbroken gauge group is $U(2)$. We also get the additional amplitudes involving the massive vectors and massive tensors 
\begin{eqnarray}
{\cal M}_3\big( 1 \overline B,2 B,3 \sugA^{i+3} \big) &=&  {i \over 2} \sigma^{i+2}  \abra{1}\aket{2} \,  \abra{2} | \varepsilon_3 | \aket{1} \ , \nn \\ 
{\cal M}_3\big( 1 \overline B,2 B,3 \sugA^1 \big) &=&  {i \over 2}   \abra{1}\aket{2} \,  \abra{2} | \varepsilon_3 | \aket{1} \ .
\end{eqnarray}
These amplitudes make explicit that the massive tensors transform in the $\bf 2$ and $\bar {\bf 2}$ representations of $U(2)$ with opposite charges.  
The amplitude 
\begin{equation}
{\cal M}_3\big( 1 \overline{W},2 W,3 \sugA^{0} \big) =  \sqrt{2} m   \langle  \overline W W A \rangle_{\overline{W} \cdot F \cdot {W} } \ 
\end{equation}
shows that massive vectors have a field-strength coupling with the massless vector obtained from the double copy of two gluons. This coupling reveals the appearance of the central charge in the Heisenberg group ${\cal H}_5$.

Finally, we introduce the additional complex gauge-theory scalar  $z = {1 \over \sqrt{2}} (\phi^1 - i \phi^2) $ and denote by $\sugA^z_\mu$ the corresponding supergravity vector from its double copy. We have the nonzero amplitude 
\begin{equation}
{\cal M}_3\big( 1 \overline B,2 W,3 \sugA^{z} \big) =  {i \over 2}
 \abra{1} \sket{2} \abra{2} | { \varepsilon}_3 | \aket{1} \ ,
\end{equation}
which indicates that this is an exotic gauging with $C_{ijM}\neq 0$. These gaugings are usually considered a very special case, but appear to emerge very naturally from the double-copy framework.

 It is worth giving more details about this gauging at the level of the algebra. 
The maximal subgroup of the global symmetry group $SU^*(6)$ is $SO^*(6)=SU(3,1)$ whose Lie algebra  admits a three grading of the form $SU(3,1)=\bar{3}  \oplus U(3) \oplus 3$, which naturally leads to a five-grading of the form  ${\cal G}_{-2} \oplus {\cal G}_{-1} \oplus {\cal G}_0 \oplus {\cal G}_1 \oplus {\cal G}_2 $ by decomposing $SU(3)$with respect to $SU(2) \oplus U(1)$. The various terms in this decomposition are
\begin{eqnarray}
&& {\cal G}_{-2} = \bar {\bf 2}^{-2}_{-2} \ , \no \\
&& {\cal G}_{-1} =  \bar {\bf 2}^{-2}_{-1} \oplus 1^0_{-1}  , \no \\
&& {\cal G}_0 = U(1)_Q \oplus U(1)_{\tilde Q} \oplus SU(2) \ ,  \\
&& {\cal G}_1 = {\bf 2}^2_1 \oplus 1^0_1  , \no \\
&& {\cal G}_2 = {\bf 2}_{2}^2  \nn   \ .
\end{eqnarray}
A linear combination $Q$  of the two U(1) generators determines the five-grading and another linear combination $\tilde Q$ assigns charges $\pm 2$ to the doublets of $SU(2) $ in grade $\pm 1$ and $\pm 2$ subspaces. The other generators have vanishing $\tilde Q$ charges.  The $U(1)_{\tilde Q}$ charges are indicated as a superscript and $U(1)_Q$ charges are indicated as subscripts.

Taking the pp-wave limit of $SU(3,1)$  leads to the algebra 
$SU(3)\ltimes \mathcal{H}_7$ with the $U(1)$ generator that gives the 3-grading becoming a central charge of $\mathcal{H}_7$ \cite{Fernando:2004jt}. 
The double-copy construction  given in Table 7 corresponds to gauging the $(SU(2) \oplus U(1)_{\tilde Q}) \ltimes \mathcal{H}_5  $ subalgebra of $SU(3)\ltimes \mathcal{H}_7$ with correct assignment of ${\tilde Q}$-charges.  
The four $SU(2) \oplus U(1)_{\tilde Q}$ grade-zero generators become gauged and correspond to double-copy fields realized as scalar times vector or vector times scalar.  Four grade-one generators transforming in the fundamental and antifundamental of $ SU(2) \oplus U(1)_{\tilde Q}$ become massive vectors. The remaining two grade-$\pm 1$  generators with vanishing $\tilde Q$-charge correspond to vectors that stay massless, and have nonzero commutators with the other four grade-one generators. Finally, four tensor fields that transform as doublets of $SU(2) \oplus U(1)_{\tilde Q}$ correspond to grade-two and grade-minus-two generators.  
The construction for the quaternionic theory presented in this section can be modified by matching masses as $m_1=-\tilde m_1$ and $m_2 = \tilde m_2$, as we have done in Section \ref{sec-hybrid}. The result is a double-copy construction for a $U(1,1)\ltimes {\cal H}_5$ gauging with four massive tensors, which has a truncation reproducing the theory studied in Section \ref{sec-hybrid}.

\section{Conclusion and outlook}

In this paper, we have introduced a framework for considering non-compact and solvable-group gaugings, applied to $\cN=2$  YME supergravity theories, using the double-copy construction for suitably chosen gauge theories. 
This framework is analogous to the one employed for gauged $\cN=8$ supergravity in \rcite{Chiodaroli:2018dbu}, with central ingredients being gauge theories coupled to massive fermions, or matter multiplets, that satisfy C/K duality.
This guarantees that, even though the matching with supergravity parameters was done at the three-point level, valid higher-point gravitational amplitudes automatically follow.  
Our construction covers
both unbroken and spontaneously broken supergravity gauge groups.

While we 
obtained several examples of non-compact gaugings, it is clear that further generalizations should be possible. Most gaugings we have constructed appear to share a common 
feature that the supergravity gauge group is the semi-direct product of a compact group with several nilpotent generators. Inspection of the amplitudes reveals that the gauge algebra includes an Heisenberg subalgebra due to the appearance of a central charge. Interestingly, this is almost a generic feature of the double copy in the presence of massive gauge-theory fermions, in the sense that the vector field corresponding to the central element of the algebra is the special supergravity vector originating from the double copy of two gauge-theory gluons. 
This structure is consistent with the Scherk-Schwarz reductions from a six dimensional theory in the presence of masses \cite{Ferrara:1996wv,Ferrara:1997gh}. An open question is the realization of more general gaugings as double copies. 
The simplest such example 
is a $U(2,1)$ gauging of the complex magical theory. While our framework gives the right field content for this theory, the  $U(2) \ltimes \mathcal{H}_5$ gauging we have obtained can be seen as a particular contraction (or pp-wave limit) of a more general $U(2,1)$ gauging. We should also mention that the procedure outlined in this paper can be straightforwardly applied to homogeneous supergravities beyond the magical and generic Jordan family of theories, including in particular the generic non-Jordan family, for which we get $U(1) \ltimes {\cal H}_{2P+1}$ gaugings.

It is worth emphasizing that the five-dimensional spinor-helicity formalism of  \rcite{Chiodaroli:2021eug} has been essential for a streamlined treatment of the theories discussed in this paper. 
Moreover, the use of complex representations for matter fields represents a departure from the  massless constructions for magical and homogeneous theories in \rcite{Chiodaroli:2015wal}, in which matter fields were in pseudo-real representations.
Complex representations are indeed more natural for massive fields since, in 
many cases, masses can be understood as compact higher-dimensional momenta and their 
signs differentiates between a representation and its conjugate.
Notably, the tools we have presented here do not provide a direct construction of gaugings of the real and octonionic magical theories. To obtain these theories, an additional projection on the output of the double copy appears to be required. Without such a projection one simply obtains homogeneous supergravities with  $q=1, P=2$  and  $q=8, P=2$. 
In the latter case, the simplest example of nontrivial gauging  still has a single $SU(2)$ factor and is very similar to the case discussed in the previous sections.
Referring to the explicit gamma matrices outlined in the appendix and without loss of generality, we use directions 11, 12 and 13 for the trilinear couplings. The resulting mass matrix acting on the $SO(9)$ spinor indices is already diagonal,
\begin{equation}
M =  {\lambda \over 4}  \sigma^3 \otimes \sigma^3 \otimes \mathbb{I}_2 \otimes \sigma^3 \ .
\end{equation}
Because of this gauging, the global symmetry is broken to $SO(6) \subset SO(9)$. However, the fermionic double copies give a total of 16 massive tensors and 16 massive vectors, $W_{ia}, \overline{W}^{ia},B_{ia}, \bar B^{ia}$, where $i$ is an internal $SU(2)$ index, and $a$ is an internal $SU(4)\cong SO(6)$ index. 
Hence, this gives  a gauging of the form 
\begin{equation}
U(2) \ltimes \mathcal{H}_{17}
\end{equation}
for the $q=8, P=2$ homogeneous theory.
Ultimately, in order to recover the octonionic magical theory, an additional reality condition needs to be imposed on the output of the double copy, which would take the schematic form 
\begin{equation}
 \overline{W}^{ia} = \epsilon^{ij} C^{ab}  W_{jb} \ ,
\end{equation}
where $C^{ab}$ acts as a charge-conjugation matrix for the $SU(4)$ index. Realizing these conditions requires a careful analysis of the amplitudes to ensure that the relevant interactions are not projected out.  Ultimately, whether a double copy can be formulated for gaugings of the octonionic theory is an important open question.

A related and well known problem is how to obtain gauge groups with large rank in the double-copy framework, since at present the only known mechanism to turn on non-abelian gauge interactions in the supergravity theory is the addition of trilinear scalar couplings in one of the gauge theories, and this can affect only supergravity vectors obtained as the double copy of a gauge-theory scalar with a gauge-theory vector. Despite considerable progress in extending the double copy, a solution to this problem remains elusive.

Given the need for possible extensions of this framework, it is natural to consider 
further generalizations of the gauge-theory factors that enter the double-copy construction. Here we have used field-theory orbifolds as a tool to generate theories that obey C/K duality starting from a simpler parent theory, as originally introduced in \rcite{Chiodaroli:2013upa}. An important open question is whether there exist more general procedures to truncate theories preserving C/K duality. 
In particular, it is currently unclear whether all consistent truncations of supergravity theories need to be symmetry-based, or if all truncations must be realized as symmetry-based projections of the gauge-theory factors entering the double-copy construction. A possible example of the latter are the non-factorizable orbifolds of ${\cal N}=8$ supergravity discussed in ref. \cite{Carrasco:2012ca}.

Finally, the theories studied in this work need to be revisited, and better understood, in six dimensions. 
Six-dimensional versions of the double copy have received relatively little attention thus far, and basic questions about the construction are still to be investigated. It would be particularly interesting to describe the theories studied in~\rcites{Ferrara:1997gh,Ferrara:1996wv} from the double-copy perspective.
 From a higher-dimensional perspective, gauging algebras with central charges seems a particularly natural choice. We plan to return on some of these issues in future work.

\section*{Acknowledgments}

This research is supported in part by the Knut and Alice Wallenberg Foundation under grants KAW 2018.0116 ({\it From Scattering Amplitudes to Gravitational Waves}) and KAW 2018.0162. The work of MC is also supported by the Swedish Research Council under grant 2019-05283. MG would like to thank the Stanford Institute for Theoretical Physics for its hospitality where part of this work was carried out. RR is supported by the US Department of Energy under Grant No. DE-SC00019066.

\newpage

\appendix

\section{Spinor-helicity formalism in five dimensions \label{AppA}}

In this appendix, we give a summary of the five-dimensional spinor-helicity formalism, referring the reader to \rcite{Chiodaroli:2022ssi} for a complete treatment. 

Amplitudes are given using a spacetime metric with  mostly-minus signature. Because of the isomorphism $SO(4,1)\sim USp(2,2) $, massless momenta contracted with five-dimensional gamma matrices give objects with two $USp(2,2)$ fundamental indices,
\be
p_{AB}=p_\mu \Gmat_{AB}^\mu \equiv -p_\mu (\Gmat^\mu)^{\ C}_A \Omega_{CB} =
-(\slash \hskip-2.4mm p \, \Omega)_{AB}\ ,
\label{pLowerAB}
\ee
where $A,B$ are fundamental $USp(2,2)$ indices, and the symplectic matrix $\Omega$ has been used to lower the second index (since gamma matrices naturally have one low and one high spinor index). In our conventions, we take $\Omega$ to be in the form
\be \label{OmegaDef}
\Omega_{AB}= \left(\begin{matrix}
\epsilon_{\alpha \beta} & 0 \\
0&  - \epsilon^{\dot \alpha \dot \beta}   \\
\end{matrix}\right)= \Omega^{BA}\, .
\ee
The indices $A,B$ are decomposed as $A=\alpha\oplus \dot\alpha$. 
The two-dimensional Levi-Civita symbol is normalized as $ \epsilon^{12}=\epsilon_{21}=1$. 

The gamma matrices with lowered indices, $\Gmat^\mu_{AB}= \Omega_{BC} (\Gmat^\mu)_A^{\ C}$, are antisymmetric and $\Omega$-traceless. Aside from the Clifford algebra identity $\{ \Gmat^\mu , \Gmat^\nu\} = 2 \eta^{\mu\nu}$, gamma matrices obey  the extra identity
\be
 (\Gmat^\mu)_A^{\ B} (\Gmat_\mu)_C^{\ D} =- 2 \Omega_{AC} \Omega^{BD}+  2\delta_{A}^D \delta_{C}^B - \delta_{A}^B \delta_{C}^D\,.
\ee
Since  $p_{AB}$ has rank two, it can be written as 
\be
p_{AB}=|p_a\rangle_{A} | p^a \rangle_{B}
~~~~~~~{\rm or}~~~~~~
{\slash \hskip-2.4mm p} {}_A{}^B\equiv p_{A}{}^{B}=|p_a\rangle_{A} \langle p^a |^{B} \ ,
\label{pmassless}
\ee
in terms of $USp(2,2)$ spinors which are labelled by extra $SU(2)$ little-group indices $a,b,\ldots$ which are also lowered and raised through left-multiplication with the Levi-Civita symbol $\epsilon^{ab}=\epsilon_{ba}$. Here we also define $\langle p^a  | {}^B=  \Omega^{BA} | p^a \rangle_{A}$. The inner product between two spinors obtained from the same massless momentum gives,
\be \label{OSconstraint}
\langle p^a  | p^b\rangle \equiv  \Omega^{BA} | p^a \rangle_{A}   | p^b \rangle_{B} =0\, .
\ee
Which is equivalent to requiring that $| p_a \rangle_{B} $ satisfies the massless Dirac equation.

To extend this construction to massive momenta in five spacetime dimensions we  split a massive momentum $p$ into two massless momenta $k$ and $q$,
\be \label{massiveMom}
p^\mu = k^\mu+ m^2 q^\mu\,,
\ee
where $q^\mu$ is null and obeys the constraint 
\begin{equation}
2 p \cdot q =2 k \cdot q=1 \ .\label{qcons}
\end{equation}
After using \eqn{pmassless} for the massless momenta, we get a corresponding expression in terms of the $SU(2)\times SU(2)$ little-group indices $a,b$ and $\dot a, \dot b$,
\be
 p_{AB} = |k_a \rangle_A  |k^a \rangle_B +  m^2 |q_a \rangle_A  |q^a\rangle_B  = {1 \over 2} 
 | {\bm p}_a \rangle_A    | {\bm p}^a \rangle_B  + 
 {1 \over 2} 
 |{\bm p}_{\dot a}]_A    |{\bm p}^{\dot a}]_B\,.
\ee
In this expression, we have defined two new massive spinors, each carrying a fundamental index with respect to one of the little-group factors,
\bea
|{\bm p}^a\rangle_A  &=& 
|k^a \rangle_A +  m  |q^a \rangle_A \,, \nn \\ 
|{\bm p}^{\dot a} ]_A  &=& |k^{\dot a} \rangle_A - m  |q^{\dot a} \rangle_A \,.
\label{massivespinors}
\eea
Following standard notation, spinors corresponding to massive momenta are in boldface.
We furthermore demand that the reference spinor $| q^a \rangle $ satisfies the constraint
\be \label{q_constraint}
\langle k^a  | q^b\rangle =  \epsilon^{ab} \,,
\ee 
which yields automatically eq. (\ref{qcons}). Moreover,  \eqn{q_constraint} implies the following identities,
\be
\langle {\bm p}^a  | {\bm p}^b \rangle = 2 
m \epsilon^{ab}\,,~~~  \qquad 
[ {\bm p}^{\dot a} | {\bm p}^{\dot b} ] = - 2 
m \epsilon^{\dot a \dot b}  \,,~~~  \qquad 
\langle {\bm p}^{a} |  {\bm p}^{\dot b} ] =0\,,
\ee
as well as the fact that the massive spinors are solutions of the Dirac equations with opposite signs for the mass,
\bea
p^{A B}|  {\bm p}^a \rangle_B &=& -m \langle {\bm p}^a|^A\,, \nn \\
p^{AB}|  {\bm p}^{\dot a}]_B &=& m \, [{\bm p}^{\dot a} |^A\, .
\eea
We proceed to list some useful identities. 
In the above spinor notation, the $USp(2,2)$ identity operator can be written as
\be
| q_a\rangle_A  \langle k^a  |^B +| k_a\rangle_A   \langle q^a  |^B =\delta_{A}^B \ .
\ee
The massive spinors obey the following completeness relations,
\bea \label{spinorCompl}
&& | {\bm p}_a\rangle_A  \langle {\bm p}^a  |^B = 
p_{A}^{\ B} + m\, \delta_{A}^B\equiv 2 m ({\cal P}^+)_{A}^{\ B}\,, \nn \\
&& | {\bm p}_{\dot a}]_A  [ {\bm p}^{\dot a}|^B = 
p_{A}^{\ B} - m\, \delta_{A}^B  \equiv 2 m ({\cal P}^-)_{A}^{\ B}\,,
\eea
where ${\cal P}^{\pm} $ are projectors satisfying  ${\cal P}^{\pm} {\cal P}^{\pm}= \pm {\cal P}^{\pm}$ and ${\cal P}^{\pm} {\cal P}^{\mp}= 0$.

In giving explicit expressions for amplitudes, it is sometimes convenient to adopt a short-hand notation based on introducing auxiliary variables to contract free little-group indices,
\be
|\,\ket{i\,} \equiv | k^a_i \rangle \auxu_{ia}  
\,,~~~~ \qquad
|\,\aket{i\,} \equiv | {\bm p}^a_i \rangle \auxu_{ia}
\, ,~~~~ \qquad 
|\,\sket{i\,} \equiv | {\bm p}^{\dot a}_i ] \auxv_{i{\dot a}} \ , \label{compactnotation}
\ee
where the index $i$ is the particle label.

We continue by giving expressions for 5D polarization vectors using our spinor variables. 
Massless vectors polarizations carry two symmetric   $SU(2)$ little-group indices and have expressions
\be
\varepsilon^\mu_{a  b} (k,q) =  \frac{ \langle k_{(a}| \Gmat^\mu   | q_{b)} \rangle}{\sqrt{2}}=  -\frac{ \langle q_{(a}| \Gmat^\mu   | k_{b)} \rangle}{\sqrt{2}}\,.
\label{masslesspol}
\ee
Massive vectors have the following polarizations:
\be
\varepsilon^\mu_{a  \dot a} (p) =  -\frac{ \langle {\bm p}_{a}| \Gmat^\mu   | {\bm p}_{{\dot a}}]}{2\sqrt{2}m}\, .
\label{massivepol2}
\ee
As explained above , free little-group indices can be dressed with auxiliary  variables $\auxu^a$, $\auxv^{\dot a}$ as follows,
\begin{equation}
\varepsilon^\mu_i = \varepsilon^\mu_{ab}(k_i,q_i)  \auxu^{a}_i \auxu^{b}_i \ , \qquad \qquad
\bep^\mu_i = \varepsilon^\mu_{a \dot b}(p_i)  \auxu^{a}_i \auxv^{\dot b}_i \ .
\end{equation}
Both massive and massless polarizations are by definition transverse and obey little-group and Minkowski space completeness relations (see  \rcite{Chiodaroli:2022ssi} for details). 

In five dimensions, there exist an additional kind of asymptotic states which correspond to massive tensors obeying self-duality (or anti-self-duality) conditions \cite{Townsend:1983xs}.  
Their polarizations 
for  massive self-dual and anti-self-dual tensors  are
\begin{align}
\varepsilon^{\mu\nu}_{a  b} (p) =  \frac{ \langle {\bm p}_{a}| \Gmat^{\mu \nu }  | {\bm p}_{b} \rangle}{4\sqrt{2}m}\ , 
\qquad \qquad
\varepsilon^{\mu\nu}_{\dot a  \dot b} (p) =  \frac{ [{\bm p}_{\dot a}| \Gmat^{\mu \nu }  | {\bm p}_{\dot b} ]}{4\sqrt{2}m}\, .
\label{poltensors2}
\end{align}

As for the vector polarizations, these tensor polarizations are transverse and satisfy a little-group completeness relation for each $SU(2)$ factor, as well as the Minkowski-space completeness relation. The polarization \eqref{poltensors} satisfy 
the (anti)self-duality relations,
\bea
p^\rho    \epsilon_{\mu\nu \rho \sigma \lambda} \,\varepsilon^{\mu\nu}_{a  b}(p) &=& - 2m\, \varepsilon_{\sigma \lambda,a  b}(p)\,, \nn \\
p^\rho    \epsilon_{\mu\nu \rho \sigma \lambda} \,\varepsilon^{\mu\nu}_{\dot a  \dot b}(p) &=& 2m\, \varepsilon_{\sigma \lambda, \dot a  \dot b}(p)\,.
\eea

Finally, external spinor polarizations obeying the standard on-shell conditions 
\begin{equation} \qquad (\cancel{p} - m ) u(p) =   \ , \qquad \bar v(p) (\cancel{p} + m ) = 0  \ ,
	\end{equation} 
are taken as 
\begin{equation}
u(p)= |{\bf p}^{a} \rangle \ , \qquad
\bar v(p)= \langle {\bf p}^{a}| . \qquad
\end{equation}
in the supersymmetric gauge-theory and 
\begin{equation}
u(p)= |{\bf p}^{\dot a}], \qquad
\bar v(p)= [ {\bf p}^{\dot a}| . \qquad
\end{equation}
in the non-supersymmetric theory. The fermionic propagator is 
\be i {\cancel{p} + m_i \over p^2 - m_i^2} \ . \ee
Note that spinor momenta in the amplitudes we calculate are taken as incoming.

This choice of spinor polarizations obeys reality conditions of the form 
\begin{equation}
(|{\bm p}^{a} \rangle_{R})^* = B E_{ab} |{\bm p}^{ b} \rangle_{\bar R}        \end{equation}
where $B=C \gamma^0 $ is the $B$ matrix and $E$ is a unit-determinant matrix acting on the little group indices as explained in  \rcite{Chiodaroli:2022ssi}, where the reader can also find explicit representations for the spinors and the matrix $E$ in terms of five-dimensional momenta under consideration. In particular, $E$ does not need to be the two-dimensional Levi-Civita symbol.

\section{Gamma matrices\label{appB}}

Here we collect explicit representations for Gamma matrices which are employed for the construction of the theories under consideration along the lines outlined in \rcite{Chiodaroli:2015wal}. The main variant with respect to the construction in  \rcite{Chiodaroli:2015wal} is that we work directly in five dimensions. More specifically, the theories under consideration can all be uplifted to an even higher dimension; in this case, it is natural to look for Gamma matrices such that a higher-dimensional spinor can be written as
\begin{equation}
    \chi_{\alpha \zeta a} \ ,
\end{equation}
where $\alpha$ is a five dimensional spinor index, $a$ is a spinor index for the compact $SO(D-5)$ group and $\xi=1,2$ is an additional spinor index. Denoting as $\gamma_\mu$ the spacetime gamma matrices and as $\tilde \gamma_i$ the gamma matrices for the compact $SO(D-5)$, we can write $D$-dimensional gamma matrices as
\begin{eqnarray}
 \Gamma^\mu &=& \gamma^\mu \otimes \sigma^1 \otimes \mathbb{I} \ , \, \qquad \mu = 0, \ldots, 4  \ , \nn \\
 \Gamma^{4+j} &=& i \, \mathbb{I} \otimes \sigma^2 \otimes \tilde \gamma^j \ , \qquad j = 1, \ldots, D-5  \ .  \end{eqnarray}
This choice leads naturally to a chirality matrix in $D$-dimensions $\Gamma_* = \mathbb{I} \otimes \sigma^3 \otimes \mathbb{I}$ for all $D$.
Since we consider irreducible spinors in $D$ dimensions, and $D$ is even for all cases explicitly studied, we can fix the additional index $\xi\equiv 1$ by taking $D$-dimensional spinors of positive chirality; this leads to a formalism where reduction to five dimensions is completely straightforward.

More explicitly, we take the gamma matrices in five dimensions as: 
\begin{eqnarray}
\gamma^0 =  \sigma^1 \otimes 1 \ , &\qquad & \gamma^3 = i  \sigma^2 \otimes \sigma^3   \ , \nn \\
\gamma^1 = i \sigma^2 \otimes \sigma^1 \ , &&  \gamma^4 = i  \sigma^3 \otimes 1  \ . \\
\gamma^2 = i \sigma^2 \otimes \sigma^2 \ , \nn
\end{eqnarray}
The charge conjugation matrix in five dimensions is 
$C_5 = i 1 \otimes \sigma^2 = \Omega$, while the $B$ matrix is $B_5 = \sigma^1 \otimes \sigma^2$ and obeys $B_5^*B_5=-1$, $B_5 \gamma^\mu B_5^{-1} = +(\gamma^\mu)^*.$
The physical masses are related to the $D$-dimensional mass matrix $M$ as 
\begin{equation}
 M = - i \mathbb{I} \otimes \sigma^2 \otimes \tilde M \ , \label{physicalmasses}
\end{equation}
where $\tilde M$ is a matrix with spinor indices corresponding to the compact dimension that has the physical masses of the fermions as eigenvalues. 

In the construction for massless, theories, it is natural to use pseudo-real gauge-group representations and impose the reality condition
\begin{equation}
\overline{\Psi} = \Psi^t C  V \ ,
\end{equation}
where $V$ is an antisymmetric matrix in the gauge indices, together with the chirality condition, reducing the fermionic on-shell degrees of freedom to two (which corresponds to a half-hyper). However, when we turn on masses, it becomes more natural to consider complex gauge-group representations. This is a consequence of the fact that the most straightforward way to introduce masses is through dimensional reduction in the presence of some compact momenta, where the sign of the momentum of a massive field now differentiates between a representation and its conjugate. In the following, we study in detail Gamma matrices and spinors for the main cases discussed in the main body of this paper, namely $D=6,8,10,14$. This choice of Gamma matrices can also be used for reproducing the results of ref. \cite{Chiodaroli:2015rdg}.

{$\bm{D=6}$}. The first case that needs to be discussed is the $\cN=2$ gauge-theory factor, which can be uplifted to six dimensions. In this case, we have just one compact gamma matrix $\tilde \gamma$, which can be taken to be equal to one. The matrix $\Gamma^5$ becomes then 
\begin{eqnarray}
\Gamma^5 = i 1 \otimes 1 \otimes \sigma^2 \ . 
\end{eqnarray}
The charge conjugation matrix and B matrix in six dimensions are $C=C_5\otimes \sigma^1$, $B=B_5\otimes 1$. The combinations $C_5\gamma$ and $C \Gamma$ are all antisymmetric. We have manifestly $[\Gamma_*,B]=0$ and, because $B^*B=-1$, we can impose a reality condition if we use pseudo-real gauge group representations.

{$\bm{D=8}$}. This choice is appropriate for the non-supersymmetric theory entering the construction for the magical complex theory. We use the following Gamma matrices for the compact directions:
\begin{eqnarray}
\Gamma^5 &=& i \mathbb{I} \otimes \sigma^2 \otimes \sigma^1 \ , \nn \\ 
\Gamma^6 &=& i \mathbb{I} \otimes \sigma^2 \otimes \sigma^2  \ ,  \\ 
\Gamma^7 &=& i \mathbb{I} \otimes \sigma^2 \otimes \sigma^3  \ . \nn
\end{eqnarray}
The charge conjugation and $B$ matrix become 
\begin{equation}
C = C_5 \otimes  \sigma^3 \otimes \sigma^2  \ , \qquad  B = B_5 \otimes  \sigma^2 \otimes \sigma^2  .
\end{equation}
In this case, the combinations $C\Gamma$ are all symmetric. Since $[\Gamma_*,B]\neq 0$, we cannot impose a reality condition if we want to work with chiral spinors.

{$\bm{D=10}$}. This choice gives the non-supersymmetric gauge theory in the construction for the quaternionic theory. We use the following Gamma matrices: 
\begin{eqnarray}
\Gamma^5 &=& i \mathbb{I} \otimes \sigma^2 \otimes \sigma^1 \otimes 1  \ , \nn \\ 
\Gamma^6 &=& i \mathbb{I} \otimes \sigma^2 \otimes \sigma^2 \otimes 1  \ , \nn \\ 
\Gamma^7 &=& i \mathbb{I} \otimes \sigma^2 \otimes \sigma^3 \otimes \sigma^1 \ , \\ 
\Gamma^8 &=& i \mathbb{I} \otimes \sigma^2 \otimes \sigma^3 \otimes \sigma^2 \ ,\nn  \\ 
\Gamma^9 &=& i \mathbb{I} \otimes \sigma^2 \otimes \sigma^3 \otimes \sigma^3 \ . \nn
\end{eqnarray}
The charge conjugation and $B$ matrix have expressions 
\begin{equation}
C = C_5 \otimes  \sigma^1 \otimes \sigma^1 \otimes \sigma^2 \ , \qquad  B = B_5 \otimes  1 \otimes \sigma^1 \otimes \sigma^2  \ ,
\end{equation}
With this choice $C \Gamma$ are all symmetric, and this is sufficient if we want to work with complex gauge-group representations.
In case one wants to use a pseudo-real representation,  the above matrices are augmented by one last 2 by 2 factor that is interpreted as acting on the flavor indices. The charge conjugation and $B$ matrix become 
\begin{equation}
C = C_5 \otimes  \sigma^1 \otimes \sigma^1 \otimes \sigma^2 \otimes \sigma^2_F \ , \qquad  B = B_5 \otimes  1 \otimes \sigma^1 \otimes \sigma^2 \otimes \sigma^2_F \ ,
\end{equation}
where the last factor acts on the extra flavor index. Taking the flavor index into account, the combinations $C\Gamma$ are antisymmetric and $\tilde C \tilde \gamma$ are symmetric, as desired.

{$\bm{D=14}$}. Finally, the extra Gamma matrices for the octonionic construction are: \begin{eqnarray}
\Gamma^5 &=& i\, \mathbb{I} \otimes \sigma^2 \otimes \sigma^1 \otimes 1 \otimes 1 \otimes 1 \ ,\nn \\ 
\Gamma^6 &=& i\, \mathbb{I} \otimes \sigma^2 \otimes \sigma^2 \otimes 1 \otimes 1 \otimes 1 \ , \nn\\ 
\Gamma^7 &=& i\, \mathbb{I} \otimes \sigma^2 \otimes \sigma^3 \otimes \sigma^1 \otimes 1 \otimes 1 \ ,\nn \\ 
\Gamma^8 &=& i\, \mathbb{I} \otimes \sigma^2 \otimes \sigma^3 \otimes \sigma^2 \otimes 1 \otimes 1 \ , \nn \\ 
\Gamma^9 &=& i\, \mathbb{I} \otimes \sigma^2 \otimes \sigma^3 \otimes \sigma^3 \otimes \sigma^1 \otimes 1 \ ,  \\ 
\Gamma^{10} &=& i\, \mathbb{I} \otimes \sigma^2 \otimes \sigma^3 \otimes \sigma^3 \otimes \sigma^2 \otimes 1 \ , \nn \\ 
\Gamma^{11} &=& i\, \mathbb{I} \otimes \sigma^2 \otimes \sigma^3 \otimes \sigma^3 \otimes \sigma^3 \otimes \sigma^1 \ , \nn \\ 
\Gamma^{12} &=& i\, \mathbb{I} \otimes \sigma^2 \otimes \sigma^3 \otimes \sigma^3 \otimes \sigma^3 \otimes \sigma^2 \ ,\nn \\ 
\Gamma^{13} &=& i\, \mathbb{I} \otimes \sigma^2 \otimes \sigma^3 \otimes \sigma^3 \otimes \sigma^3 \otimes \sigma^3  \ . \nn
\end{eqnarray}
In this case, the charge conjugation and $B$ matrix are 
\begin{equation}
C = C_5 \otimes  \sigma^1 \otimes \sigma^1 \otimes \sigma^2 \otimes \sigma^1 \otimes \sigma^2 \ , \qquad  B = B_5 \otimes  1 \otimes \sigma^1 \otimes \sigma^2 \otimes \sigma^1 \otimes \sigma^2 
\end{equation}
The combinations $C \Gamma$ are antisymmetric, while the combinations $\tilde C \tilde \gamma$ are symmetric in the spinor indices. One has also $B^*B=-1$, indicating that one needs to put the gauge fields in a pseudo-real representation (in case of the massless construction), or impose an additional condition after the double copy. 
\newpage

\bibliographystyle{JHEP}
\bibliography{litCGJR2021}

\end{document}